\documentclass[a4paper,12pt]{article}
\pdfoutput=1
\usepackage{amssymb,amsmath}
\usepackage{mathrsfs}
\usepackage{bbm}
\usepackage{graphicx,subfigure}
\newlength{\dinwidth}
\newlength{\dinmargin}
\begin{document}

\title{\bf Anomalous $tq\gamma$ coupling effects in exclusive radiative B-meson decays}
\bigskip

\author{Xin-Qiang Li$^{1,2}$, Ya-Dong Yang$^{3,4}$ and Xing-Bo Yuan$^{3}$\\
{ $^1$\small Department of Physics, Henan Normal University, Xinxiang, Henan 453007, P.~R. China}\\
{ $^2$\small IFIC, Universitat de Val\`encia-CSIC, Apt. Correus 22085, E-46071 Val\`encia, Spain}\\
{ $^3$\small Institute of Particle Physics, Huazhong Normal University, Wuhan, Hubei 430079, P.~R. China}\\
{ $^4$\small Key Laboratory of Quark \& Lepton Physics, Ministry of Education, Huazhong Normal University}\\
{\small Wuhan, Hubei, 430079, P.~R. China}}

\date{}
\maketitle
\bigskip \bigskip
\maketitle
\vspace{-1.5cm}

\begin{abstract}
{\noindent}
The top-quark FCNC processes will be searched for at the CERN LHC, which are correlated with the B-meson decays. In this paper, we study the effects of top-quark anomalous interactions $tq\gamma$ in the exclusive radiative $B\to K^*\gamma$ and $B\to\rho\gamma$ decays. With the current experimental data of the branching ratios, the direct CP and the isospin asymmetries, bounds on the coupling $\kappa_{tcR}^{\gamma}$ from $B\to K^*\gamma$ and $\kappa_{tuR}^{\gamma}$ from $B\to \rho\gamma$ decays are derived, respectively. The bound on $|\kappa_{tcR}^{\gamma}|$ from ${\mathcal B}(B\to K^{*}\gamma)$ is generally compatible with that from ${\mathcal B}(B\to X_{s}\gamma)$. However, the isospin asymmetry $\Delta(K^{*}\gamma)$ further restrict the phase of $\kappa_{tcR}^{\gamma}$, and the combined bound results in the upper limit, $\mathcal B(t\to c\gamma)<0.21\%$, which is lower than the CDF result. For real $\kappa_{tcR}^{\gamma}$, the upper bound on  $\mathcal B(t\to c\gamma)$ is about of the same order as the $5\sigma$ discovery potential of ATLAS with an integrated luminosity of $10~{\rm fb}^{-1}$. For $B\to\rho\gamma$ decays, the NP contribution is enhanced by a large CKM factor $|V_{ud}/V_{td}|$, and the constraint on $tu\gamma$ coupling is rather restrictive, $\mathcal B(t\to u\gamma)<1.44\times 10^{-5}$. With refined measurements to be available at the LHCb and the future super-B factories, we can get close correlations between $B\to V \gamma$ and the rare $t\to q\gamma$ decays, which will be studied directly at the LHC ATLAS and CMS.

\end{abstract}

\newpage

\section{Introduction}
\label{Sec:intro}

In the Standard Model~(SM), the flavor-changing neutral current~(FCNC) interactions, which are absent at the tree level, are highly suppressed at one loop due to the Glashow-Iliopoulos-Maiani~(GIM) mechanism~\cite{Glashow:1970gm}. Possible new physics~(NP) beyond the SM can  manifest itself by altering the expected rates of these FCNC processes. Thus, the study of FCNC interactions plays an important role in testing the SM and probing NP beyond it.

For the top quark in particular, the FCNC decays $t \to q \gamma$~(where $q$ denotes either a $c$- or a $u$-flavored quark) are expected to be exceedingly rare within the SM, with branching ratios of order of $10^{-10}$~\cite{Eilam:1990zc}. Any positive signal of these decays would  imply NP beyond the SM. These top-quark anomalous couplings could also be probed by studying the top quark production~\cite{Beneke:2000hk}. Present constraints on the FCNC couplings $tq\gamma$ come from the following two experimental bounds: the branching ratio $\mathcal B(t \to q \gamma)<3.2\%$ at $95\%$ confidence level~(C.L.) set by the CDF collaboration~\cite{Abe:1997fz},\footnote{This upper limit corresponds to $|\kappa_{tcR}^{\gamma}/{\Lambda}|<1.089~{\rm TeV}^{-1}$ in our convention for the effective Lagrangian defined by Eq.~(\ref{Lagrangian}).} and the effective FCNC coupling $\kappa_{tu\gamma}<0.174$ at $95\%$ C.L. provided by the ZEUS collaboration~\cite{Chekanov:2003yt}.\footnote{This value corresponds to $|\kappa_{tuR}^{\gamma}/{\Lambda}|<0.469~{\rm TeV}^{-1}$ in our convention, and an upper limit $\mathcal B(t \to u \gamma)<0.59\%$ at $95\%$ C.L.~\cite{Nakamura:2010zzi}.} These constraints will be improved significantly by the large top-quark sample to be available at the CERN Large Hadron Collider~(LHC), which is expected to produce $8 \times 10^6$ top quark pairs and $3 \times 10^6$ single top quark annually, even at the initial low luminosity run~($10~{\rm fb}^{-1}/{\rm year}$). In particular, both the ATLAS~\cite{Carvalho:2007yi} and the CMS~\cite{Benucci:2008zz} collaborations have presented their sensitivity to these rare top-quark decays induced by anomalous FCNC interactions, and will be able to observe $t \to q\gamma$ decays if their branching ratios could be enhanced to $\mathcal O(10^{-4})$ with $10~{\rm fb}^{-1}$ data.

However, when performing the study of anomalous top-quark interactions at the LHC and other future colliders, one should take into account constraints from some precisely measured low-energy processes where loops involving the top quarks play a crucial role. In this respect, the radiative B-meson decays induced by the FCNC transitions $b \to s\gamma$ and $b \to d\gamma$ are among the most valuable probes of anomalous top-quark couplings~\cite{FCNC-top-old,Lee:2008xr,Fox:2007in,Grzadkowski:2008mf,Yuan:2010vk}. For example, the anomalous couplings $tc\gamma$ mediating the rare radiative decays $t\to q\gamma$ could also affect the radiative $b\to s\gamma$ decays through the top-quark loops~\cite{Yuan:2010vk}. On the experimental side, both the inclusive and the exclusive $b \to s\gamma$ branching ratios are known with good accuracy~($\sim 5\%$ for $B \to K^* \gamma$ and $\sim 7\%$ for $B \to X_s \gamma$), while measurements are only available for exclusive $b \to d\gamma$ channels~\cite{Asner:2010qj}. For exclusive channels, besides the branching ratios, some other interesting observables like the CP and isospin asymmetries have also been measured~\cite{Asner:2010qj}. On the theoretical side, while the inclusive decays can be essentially calculated perturbatively, the exclusive processes are more complicated due to the non-perturbative strong interaction effects~\cite{B2Vgreview}. Besides some other methods~\cite{B2Vg-PQCD,B2Vg-SCET,Ball:2006eu,B2Vg-LCSR}, the QCD factorization~(QCDF) approach has provided a systematic framework for the treatment of exclusive radiative B-meson decays~\cite{Beneke:2001at,Bosch:2001gv,Ali:2001ez,Kagan:2001zk}. With both the experimental and the theoretical progresses, the exclusive channels are also expected to provide important constraints on the anomalous top-quark couplings~\cite{Lee:2008xr,Fox:2007in} and on various NP models~\cite{B2Vg-NP}.

In this paper, we shall perform a model-independent study of the effects of anomalous FCNC couplings $tq\gamma$ in exclusive $B\to K^*\gamma$ and $B\to\rho\gamma$ decays, and derive the corresponding constraints on their strengths. Besides the branching ratios, we shall also consider the CP and the isospin asymmetries in these decays within the QCDF framework~\cite{Beneke:2001at,Bosch:2001gv,Ali:2001ez,Kagan:2001zk}. Constraints on these anomalous couplings from the current data on these observables are then derived, and implications for rare $t \to q\gamma$ decays at the LHC are also discussed.

Our paper is organized as follows. In Sec.~\ref{Sec:NP}, we introduce the effective Lagrangian describing the anomalous interactions $tq\gamma$, and set the convention used throughout the paper. In Sec.~\ref{Sec:TF}, the theoretical framework for exclusive $B\to V\gamma$ decays within the QCDF method, and the relevant formulae for rare $t\to q\gamma$ decays mediated by these anomalous interactions are presented. How these anomalous couplings manifest themselves in these processes is also presented here. In Sec.~\ref{Sec:Na}, we give our detailed numerical results and discussions. Our conclusions are made in Sec.~\ref{Sec:Conclusions}. The relevant input parameters are collected in the appendix.

\section{Effective Lagrangian for anomalous $tq\gamma$ couplings}
\label{Sec:NP}

In most extensions of the SM, the new degrees of freedom that modify the ultraviolet behavior of the underlying theory appear only at some higher scale $\Lambda$. As long as we are only interested in processes occurring much below this scale, we can integrate out these new degrees of freedom and describe the NP effects in terms of a few higher-dimensional local operators, which are built out of the SM fields and suppressed by inverse powers of the NP scale $\Lambda$~\cite{Appelquist:1974tg}. A complete set of independent operators of dimension 5 and 6 that are consistent with the SM gauge symmetries could be found in Refs.~\cite{Grzadkowski:2010es,Buchmuller:1985jz,AguilarSaavedra:2008zc}.

Specific to the anomalous top-quark interactions $tq\gamma$, which has a magnetic dipole structure as required by gauge invariance, the relevant effective Lagrangian with dimension 5 operators can be written in a model-independent way as~\cite{Beneke:2000hk,AguilarSaavedra:2008zc,Hollik:1998vz}
\begin{equation}\label{Lagrangian}
 {\mathcal L}_5 = -e \sum_{q=u,c} \frac{\kappa^{\gamma}_{tqL}}{\Lambda}\, \bar q_R \sigma^{\mu \nu} t_L F_{\mu \nu}
 -e \sum_{q=u,c} \frac{\kappa^{\gamma}_{tqR}}{\Lambda}\, \bar q_L \sigma^{\mu \nu} t_R F_{\mu \nu} + {\rm h.c.},
\end{equation}
where $\sigma^{\mu \nu}=i\,[\gamma^{\mu},\gamma^{\nu}]/2$, $q_{R,L}=(1 \pm \gamma_5)\,q/2$ are the right- and left-handed quark fields, and $F_{\mu \nu}=\partial_{\mu}A_{\nu}-\partial_{\nu}A_{\mu}$ the photon field strength tensor. Normalized to the NP scale $\Lambda$, the coefficients $\kappa_{tqL}^\gamma$ and $\kappa_{tqR}^\gamma$ are dimensionless couplings and in general complex. The effective Lagrangian given by Eq.~(\ref{Lagrangian}) is commonly employed in phenomenological analyses related to top-quark physics~\cite{Beneke:2000hk,Carvalho:2007yi,Zhang:2008yn,Drobnak:2010wh}.

From the Dirac structure of the dimension 5 operators in Eq.~(\ref{Lagrangian}), we can see that the $tq\gamma$ vertex is induced by two independent chirality-flipped operators $m_q\,\bar q_R \sigma^{\mu \nu} t_L F_{\mu \nu}$ and $m_t\,\bar q_L \sigma^{\mu \nu} t_R F_{\mu \nu}$, where the external quark mass factors must appear in order to obtain a nonzero contribution whenever a chirality flipping $L \leftrightarrow R$ occurs. Due to the mass hierarchy $m_t \gg  m_{q}$, the effect of $m_q\,\bar q_R \sigma^{\mu \nu} t_L F_{\mu \nu}$ could be neglected unless the coupling $\kappa_{tqL}^{\gamma}$ is enhanced to be comparable to $\frac{m_t}{m_q}\,\kappa_{tqR}^{\gamma}$ by some unknown mechanism. Thus, to a good approximation, we shall focus only on the coupling $\kappa_{tqR}^{\gamma}$ in this paper~\cite{Yuan:2010vk}.

\section{Theoretical formalism}
\label{Sec:TF}

In this section, we briefly present the theoretical framework for exclusive $B\to V\gamma$ decays within the QCDF method, and the rare $t\to q\gamma$ decays mediated by anomalous $tq\gamma$ interactions. For more details, the readers are referred to Refs.~\cite{Beneke:2001at,Bosch:2001gv,Ali:2001ez,Kagan:2001zk} and \cite{Zhang:2008yn,Drobnak:2010wh}, respectively.

\subsection{$B \to V \gamma$ decays within the QCDF framework}
\label{Sec:QCDF}

\subsubsection{The effective Hamiltonian}
\label{Sec:Heff}

In the SM, the effective Hamiltonian for radiative $b\to D\gamma$~(with $D=d,s$) transitions can be written as~\cite{Beneke:2001at}
\begin{equation}\label{Eq:Heff1}
 \mathcal H_{\rm eff} = -\frac{G_F}{\sqrt{2}}\left[\lambda_t^{(D)} {\cal H}_{\rm eff}^{(t)}
 +\lambda_u^{(D)} {\cal H}_{\rm eff}^{(u)} \right]\, + \mbox{h.c.},
\end{equation}
where $\lambda_q^{(D)}=V_{qb}V_{qD}^*$, is the product of Cabibbo-Kobayashi-Maskawa~(CKM) matrix elements~\cite{CKM}, and using the unitarity relations $\lambda_u^{(D)}+\lambda_c^{(D)}+\lambda_t^{(D)}=0$, we have
\begin{eqnarray}\label{Eq:Heff2}
 \mathcal H_{\rm eff}^{(t)} &=& C_1\, \mathcal O_1^c + C_2\, \mathcal O_2^c +\sum_{i=3}^{8} C_i\, \mathcal O_i\,,\\
 \mathcal H_{\rm eff}^{(u)} &=& C_1\, (\mathcal O_1^c-\mathcal O_1^u) + C_2\, (\mathcal O_2^c-\mathcal O_2^u)\,.
\end{eqnarray}
Here we adopt the operator basis introduced by Chetyrkin, Misiak, and M\"unz~(CMM)~\cite{Chetyrkin:1996vx},
\begin{align}\label{eq:operators}
   \mathcal O^p_1 &= \bar D \gamma_\mu (1-\gamma_5) T^a p\, \bar p \gamma^\mu (1-\gamma_5) T^a b\,,
  &\mathcal O^p_2 &= \bar D \gamma_\mu (1-\gamma_5) p\, \bar p \gamma^\mu (1-\gamma_5) b\,, \nonumber \\[0.2cm]
   \mathcal O_3 &= 2 \bar D \gamma_\mu (1-\gamma_5) b\, \sum_q \bar q\gamma^\mu q\,,
  &\mathcal O_5 &= 2 \bar D \gamma_{\mu_1}\gamma_{\mu_2}\gamma_{\mu_3} (1-\gamma_5) b\, \sum_q \bar q
   \gamma^{\mu_1}\gamma^{\mu_2}\gamma^{\mu_3} q\,, \nonumber \\
   \mathcal O_4 &= 2 \bar D \gamma_\mu (1-\gamma_5) T^a b\, \sum_q \bar q\gamma^\mu T^a q\,,
  &\mathcal O_6 &= 2 \bar D \gamma_{\mu_1}\gamma_{\mu_2}\gamma_{\mu_3} (1-\gamma_5) T^a b\, \sum_q \bar q
   \gamma^{\mu_1}\gamma^{\mu_2}\gamma^{\mu_3} T^a q\,, \nonumber \\
   \mathcal O_7 &= -\frac{e}{8\pi^2}\, \bar{m}_b\, \bar D\sigma^{\mu\nu} (1+\gamma_5) b\, F_{\mu\nu}\,,
  &\mathcal O_8 &= -\frac{g_s}{8\pi^2}\, \bar{m}_b\, \bar D\sigma^{\mu\nu}T^a (1+\gamma_5) b\, G^a_{\mu\nu}\,,
\end{align}
where $T^a$ are $SU(3)_C$ generators and $\bar m_b$ denotes the b-quark mass in the $\overline{\rm MS}$ scheme. The corresponding Wilson coefficients at the lower scale $\mu=m_b$ can be calculated perturbatively~\cite{Chetyrkin:1996vx,WC1,WC2}, and their numerical values at the leading-logarithmic~(LL) and the next-to-leading-logarithmic~(NLL) order are collected in Table~\ref{tab:WC}.

%%%%%%%%%%%%%%%%%%%%%%%%%%%%%%%%%%%%%%%%%%%%%%%%%%%%%%%%%%%%%%%%%%%
\begin{table}[t]
\begin{center}
\caption{\label{tab:WC} \small Wilson coefficients at the scale $\mu=4.45~{\rm GeV}$ in the LL and the NLL order, using  two-loop running for $\alpha_s$ with the input parameters listed in the appendix.}
\vspace{0.2cm}
\doublerulesep 0.8pt \tabcolsep 0.11in
\begin{tabular}{lcccccccc}
\hline\hline
    & ${C}_1$   & ${C}_2$  & ${C}_3$   & ${C}_4$   & ${C}_5$  & ${C}_6$  & $C_7^{\rm eff}$ & $C_8^{\rm eff}$ \\
\hline
LL  & $-0.5157$ & $1.0262$ & $-0.0052$ & $-0.0696$ & $0.0005$ & $0.0010$ & $-0.3179$ & $-0.1505$ \\
NLL & $-0.3049$ & $1.0082$ & $-0.0048$ & $-0.0841$ & $0.0003$ & $0.0009$ & $-0.3078$ & $-0.1692$ \\
\hline\hline
\end{tabular}
\end{center}
\end{table}
%%%%%%%%%%%%%%%%%%%%%%%%%%%%%%%%%%%%%%%%%%%%%%%%%%%%%%%%%%%%%%%%%%%

\subsection{Factorization formula for the matrix elements}
\label{Sec:matrixelement}

Starting from the effective Hamiltonian Eqs.~(\ref{Eq:Heff1}) and (\ref{Eq:Heff2}), the matrix elements for $B\to V\gamma$ decays can be written as~\cite{Beneke:2001at,Bosch:2001gv}
\begin{equation}\label{eq:matrixelement}
 \langle V(p^{\prime},\varepsilon) \gamma(q,\eta) | {\cal H}^{(i)}_{\rm eff} | \bar B(p) \rangle
 = \frac{i\, e\, m_b}{2\,\pi^2}\, {\cal T}_\perp^{(i)}(0)\, \bigg\{\epsilon^{\mu\nu\rho\sigma}\, \eta^{\ast}_{\mu}\,
 \varepsilon^{\ast}_{\nu}\, p_{\rho} p^{\prime}_{\sigma} - i\, \Big[(\eta^{\ast} \cdot \varepsilon^{\ast})\,
 (q \cdot p^{\prime}) - (\eta^{\ast} \cdot p^{\prime})\, (\varepsilon^{\ast} \cdot q)\Big]\bigg\},
\end{equation}
where $|V\rangle$ denotes a light vector meson state and is defined as
\begin{equation}
 |V\rangle \equiv \left\{\begin{array}{l}
          |\rho^-\rangle\,, |K^{*-} \rangle\,, \qquad\,  \mbox{for $B^-$ decays}\,, \\
 -\sqrt 2 |\rho^0\rangle\,, |\bar K^{*0}\rangle\,, \qquad\, \mbox{for $\bar B^0$ decays}\,.
\end{array}\right.
\end{equation}
Using this convenient parametrization, all the dynamical information is encoded in the form factor ${\cal T}_\perp^{(i)}(0)$, which constitutes a big challenge to be determined precisely~\cite{Beneke:2001at,Bosch:2001gv,Ali:2001ez}.

In the QCDF formalism, the form factor ${\cal T}_\perp^{(i)}(0)$ could be computed in terms of heavy-to-light transition form factors and hadron light-cone distribution amplitudes~(LCDAs) at the leading power in $1/m_b$ expansion. Explicitly, it takes the following factorization formula~\cite{Beneke:2001at}
\begin{equation}\label{calT}
 {\cal T}_\perp^{(i)}(0) = T_1(0) \, C_\perp^{(i)}  + \frac{\pi^2}{N_c}\,\frac{f_B f_V^\perp}{m_B}\, \sum_{\pm}
 \int\frac{d\omega}{\omega}\, \Phi_{B,\,\pm}(\omega) \int_0^1 \!du\,\, \phi_\perp(u) \, T_{\perp,\,\pm}^{(i)}(u,\omega)\,,
\end{equation}
where $f_B$ and $\Phi_{B,\pm}$ denote the B-meson decay constant and LCDAs, and $f_V^\perp$ and $\phi_{\perp}$ the corresponding quantities of the light vector meson, respectively. The first term, expressed in terms of the tensor form factor $T_1(0)$, corresponds to the vertex corrections where the spectator quark in the B-meson does not participate in the hard process, whereas the second term incorporates the hard-scattering contributions where the spectator quark is involved. Thus, the first and the second term are usually called the ``form factor'' and the ``spectator scattering'' term, respectively.

The hard-scattering kernels $C_\perp^{(i)}$ and $T_{\perp,\,\pm}^{(i)}$ in Eq.~(\ref{calT}) can be calculated perturbatively and, up to next-to-leading order~(NLO), have the following expansions, respectively~\cite{Beneke:2001at}
\begin{align}
 C_\perp^{(i)} &= C_\perp^{(0,i)}+\frac{\alpha_s C_F}{4\pi} \,C_\perp^{(1,i)} + \ldots, \label{c11} \\[0.2em]
 T^{(i)}_{\perp,\,\pm}(u,\omega) &= T^{(0,i)}_{\perp,\,\pm}(u,\omega) + \frac{\alpha_s C_F}{4\pi} \,
 T^{(1,i)}_{\perp,\,\pm}(u,\omega) + \ldots\,, \label{t11}
\end{align}
where the QCD coupling $\alpha_s$ should be evaluated at the scale $\mu_b \simeq m_b$ in Eq.~(\ref{c11}) and at $\mu_{h} \simeq (m_b\Lambda_{\rm QCD})^{1/2}$ in Eq.~(\ref{t11}), corresponding to the typical virtualities in the two terms, respectively. Explicit expressions for the coefficients $C_\perp^{(0,i)}$, $C_\perp^{(1,i)}$ and $T^{(0,i)}_{\perp,\,\pm}$, $T^{(1,i)}_{\perp,\,\pm}$ could be found in Ref.~\cite{Beneke:2001at}.

For $B \to V \gamma$ decays, besides the leading-power contributions given by Eqs.~(\ref{c11}) and (\ref{t11}), it is also well-known that some very specific power corrections, such as the weak annihilation and exchange amplitudes in $B \to \rho\gamma$ that are enhanced by the large Wilson coefficient $C_2 \sim 1$, are often numerically important~\cite{Beneke:2001at,Bosch:2001gv,Ali:2001ez}. Furthermore, the annihilation topologies have been shown to provide the main source for the isospin asymmetry in $B \to K^* \gamma$~\cite{Kagan:2001zk}. Although being power-suppressed by $1/m_b$, these contributions are still computable within the QCDF framework. Thus, in this paper we have also included these isospin-breaking power corrections, which are denoted by $\Delta\mathcal T_\perp^{(i)}\vert_{\rm ann}$ and $\Delta\mathcal T_\perp^{(i)}\vert_{\rm hsa}$ for the weak annihilation and the hard spectator scattering, respectively. Their explicit expressions could be found in Refs.~\cite{Beneke:2001at,Feldmann:2002iw}.

For the isospin-breaking power corrections, it should be noted that an endpoint divergence is encountered in the matrix element of chromo-magnetic dipole operator $\mathcal O_8$, belonging to the term $\Delta\mathcal T_\perp^{(t)}\vert_{\rm hsa}$. Following the treatment adopted in Refs.~\cite{Kagan:2001zk,Feldmann:2002iw}, we regulate this singularity with an \textit{ad hoc} cutoff
\begin{equation}
  \int_0^1 du \rightarrow (1+\rho \, e^{i\phi}) \, \int_0^{1-\Lambda_h/m_B} du\,,
\end{equation}
and take $\Lambda_h \simeq 0.5~{\rm GeV}$, together with $0 \leq \rho\leq 1$ and $0 \leq \phi \leq 2 \pi$, to give a conservative estimation of the theoretical uncertainty related to this power correction.

\subsubsection{Observables in $B\to V\gamma$ decays}
\label{Sec:observables}

When discussing observables in $B\to V\gamma$ decays, it is more convenient to express the decay amplitudes in terms of a new quantity ${\cal C}_7^{(i)}$, defined by~\cite{Beneke:2001at}
\begin{equation} \label{CalC7}
 {\cal C}_7^{(i)} \equiv \frac{{\cal T}_{\perp}^{(i)}(0)}{T_1(0)} = \delta^{it}\,C_7^{\rm eff} + \ldots\,,
\end{equation}
where $C_7^{\rm eff}=C_7-C_3/3-4 C_4/9-20 C_5/3-80 C_6/9$, defined in the ${\rm \overline{MS}}$ scheme with fully anti-commuting $\gamma_5$, is the effective Wilson coefficient~\cite{Chetyrkin:1996vx}, and the ellipses denote the $\mathcal O(\alpha_s)$ and the sub-leading power corrections discussed in the last subsection.

In terms of the quantity ${\cal C}_7^{(i)}$, the decay rate for $\bar B\to V\gamma$ decays can then be written as~\cite{Beneke:2001at}
\begin{equation}\label{Gamma}
  \Gamma(\bar B\to V \gamma) = \frac{G_F^2}{8\pi^3}\,m_B^3\, S\left(1-\frac{m_V^2}{m_B^2}\right)^{\!3}
  \frac{\alpha_{\rm em}}{4\pi}\,m_{b}^2\,T_1(0)^2\, |\lambda_t^{(D)}{\cal C}_7^{(t)}
+ \lambda_u^{(D)}{\cal C}_7^{(u)}|^2\,,
\end{equation}
with $S=1/2$ for $\rho^0$, and $S=1$ for the other light vector mesons. In the SM, the CP-conjugated mode $\Gamma(B\to \bar{V} \gamma)$ follows from Eq.~(\ref{Gamma}) with the replacement $\lambda_i^{(D)}\to \lambda_i^{(D)\ast}$. For $b\to s$ transitions, the dominant term is $\lambda_t^{(s)}{\cal C}_7^{(t)}$, since the contributions proportional to $\lambda_u^{(s)}$ are doubly Cabibbo-suppressed; whereas for $b\to d$ transitions, where $\lambda_u^{(d)}$ is of the same order as $\lambda_t^{(d)}$, the interference term is non-negligible and can be the source of interesting CP-violating and isospin-breaking effects.

With the decay rate Eq.~(\ref{Gamma}) at hand, the interesting observables in $B\to V\gamma$ decays can be defined as follows~\cite{Beneke:2001at,Bosch:2001gv,Ali:2001ez}
\begin{itemize}
 \item the CP-averaged branching ratio
\begin{equation}\label{Br}
 \mathcal B(\bar B\to V \gamma) = \tau_B \frac{\Gamma(\bar B\to V \gamma) + \Gamma(B\to \bar{V} \gamma)}{2}\,,
\end{equation}
with $\tau_B$ being the $B$-meson lifetime.

 \item the direct CP asymmetry
\begin{equation}\label{ACP}
 \mathcal A_{CP}(V\gamma) = \frac{\Gamma(\bar B \to V \gamma) - \Gamma(B \to \bar{V} \gamma)}{
 \Gamma(\bar B \to V \gamma) + \Gamma(B \to V \gamma)}\,.
\end{equation}

 \item the isospin asymmetry
\begin{align}\label{Isospin}
 \Delta(K^*\gamma) &=\frac{\Gamma(B^0\to K^{*0}\gamma)-\Gamma(B^+\to K^{*+}\gamma)}{\Gamma(B^0\to
 K^{*0}\gamma)+\Gamma(B^+\to K^{*+}\gamma)}, \\[0.2cm]
 \Delta(\rho\gamma) &=\frac{\Gamma(B^+\to\rho\gamma)}{2\Gamma(B^0\to\rho^0\gamma)}-1\,,
\end{align}
where all decay rates are assumed to be CP-averaged.
\end{itemize}
There are five observables for $B\to K^*\gamma$ and $B\to\rho\gamma$ decays, respectively, i.e., two CP-averaged branching ratios, two direct CP asymmetries, and one isospin asymmetry. Once being measured precisely, they could be used to test the SM and to probe various NP beyond it~\cite{Lee:2008xr,Fox:2007in,B2Vg-NP}. Especially, the two isospin  asymmetries are expected to provide useful information complementary to the corresponding inclusive decay modes~\cite{B2Vg-NP}.

\subsubsection{Anomalous $tq\gamma$ coupling effects on $B\to V\gamma$ decays}

%%%%%%%%%%%%%%%%%%%%%%%%%%%%%%%%%%%%%%%%%%%%%%%%%%%%%%%%%%%%%%%%%%%
\begin{figure}[t]
\centering
\subfigure[]{\includegraphics [width=5cm]{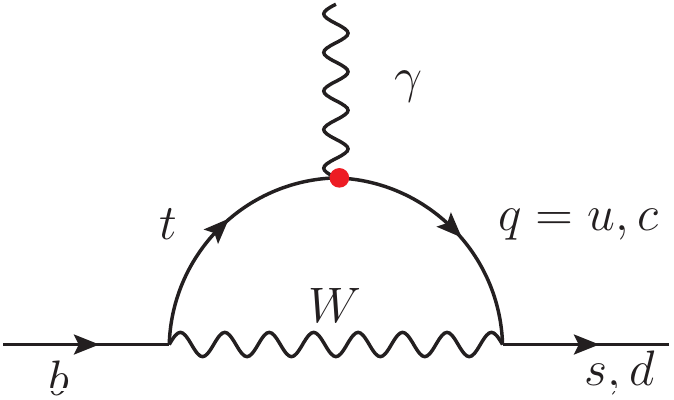}\label{fig1a}}
\subfigure[]{\includegraphics [width=5cm]{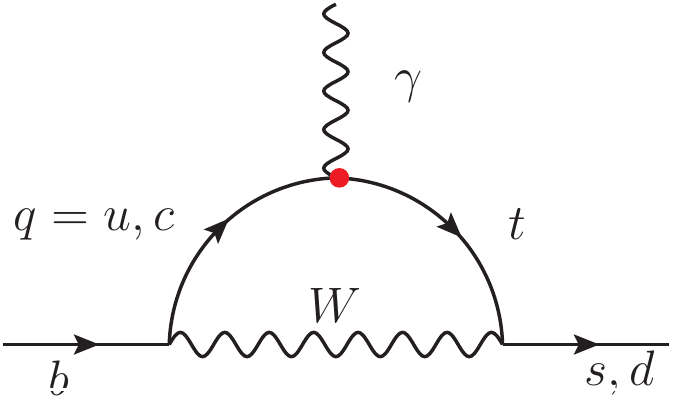}\label{fig1b}}
\subfigure[]{\includegraphics [width=5cm]{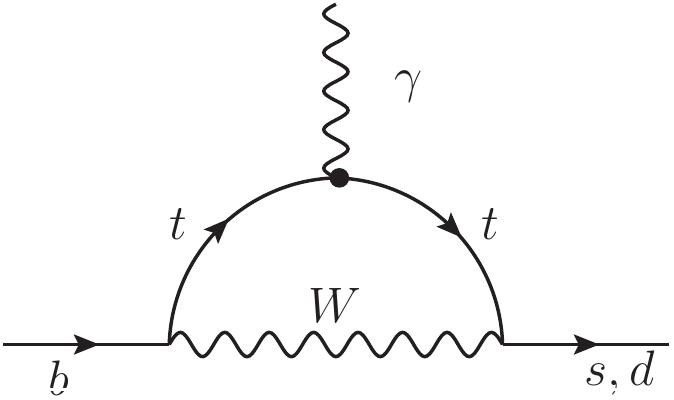}\label{fig1c}}
\caption{\small Feynman diagrams for $b \to s(d) \gamma$ transitions in the unitary gauge. (a) and (b) are mediated by the anomalous $tq\gamma $ interactions defined by Eq.~(\ref{Lagrangian}), while (c) represents a sample LO penguin diagram in the SM.}
\end{figure}
%%%%%%%%%%%%%%%%%%%%%%%%%%%%%%%%%%%%%%%%%%%%%%%%%%%%%%%%%%%%%%%%%%%

The anomalous $tq\gamma$ interactions affect $b \to D \gamma$ transitions through the two Feynman diagrams depicted in Figs.~\ref{fig1a} and \ref{fig1b}. It is interesting to note that the associated CKM factors in Figs.~\ref{fig1a} and \ref{fig1b} are $V_{tb} V_{qD}^{*}$ and $V_{qb} V_{tD}^{*}$, respectively. Since $|V_{tb} V_{qD}^{*}| \gg  |V_{qb} V_{tD}^{*}|$ for both $q=u$ and $c$ quarks, the contribution of Fig.~\ref{fig1a} would be much larger than that of Fig.~\ref{fig1b}. Furthermore, given the strengths of the couplings $tu\gamma$ and $tc\gamma$ comparable, the contribution of Fig.~\ref{fig1a} is still dominated only by one coupling, $tc\gamma$ for $b \to s\gamma$ and $tu\gamma$ for $b \to d\gamma$ respectively, because of the relations $|V_{cs}|>|V_{us}|$ and $|V_{ud}|>|V_{cd}|$. Hence we shall only consider the contribution of Fig.~\ref{fig1a} with only one anomalous coupling.

From the Feynman diagram Fig.~\ref{fig1a}, it is also observed that the large CKM factors $|V_{tb}V_{cs}|\approx 1$ and $|V_{tb}V_{ud}|\approx 1$, compared to the corresponding SM case $|V_{tb}V_{ts}|\sim \mathcal O(\lambda^2)$ and $|V_{tb}V_{td}|\sim \mathcal O(\lambda^3)$, make the transitions $b\to s\gamma$ and $b \to d\gamma$ to be very sensitive to the strengths of anomalous couplings $tc\gamma$ and $tu\gamma$, respectively. Constraint on the coupling $tc\gamma$ from the precisely measured inclusive decay $B\to X_s\gamma$ has been studied in detail by two of us~\cite{Yuan:2010vk}.

The calculation of Fig.~\ref{fig1a} can be most conveniently carried out in the unitary gauge where the pseudo-Goldstone components of the SM Higgs doublet are absent, which has been calculated in 't Hooft-Feynman gauge in Ref.~\cite{Yuan:2010vk}. It is noted that in the unitary gauge, the $g^{\alpha\beta}$ part of the $W$-boson propagator gives an ultraviolet-finite contribution, while the contribution from the $q^{\alpha}q^{\beta}/m_{W}^2$ part is ultraviolet-divergent. Following the  treatment adopted by Grzadkowski and Misiak~\cite{Grzadkowski:2008mf}, we shall apply the modified minimal subtraction~(${\rm \overline{MS}}$) scheme to  absorb  the divergences into some counterterms served by other dimension-six operators~\cite{Grzadkowski:2010es,Buchmuller:1985jz}, the ${\rm \overline{MS}}$-renormalized Wilson coefficients of which are assumed to be  negligible in comparison with  the ones considered here. With such a prescription, logarithms $\ln\frac{\mu_W}{m_{W}}$ are present in the matching coefficients.

Normalized to the effective Hamiltonian Eq.~(\ref{Eq:Heff1}) and the operator basis Eq.~(\ref{eq:operators}), the contributions of anomalous $tq\gamma$ interactions to $B\to V\gamma$ decays would result in the deviation~\cite{Yuan:2010vk}
\begin{equation}\label{C7'}
 C_7(\mu_W) \to C_7^{\prime}(\mu_W)=C_7^{\rm SM}(\mu_W)+C_7^{{\rm NP}}(\mu_W)\,,
\end{equation}
where~($x_{q}=\bar{m}_{q}(\mu_W)^2/m_{W}^2$)
\begin{align}\label{C7tcr}
 C_{7}^{{\rm NP}}(\mu_W) =\kappa_{tcR}^\gamma\, \frac{m_t}{\Lambda}\,\frac{V_{cs}^*}{V_{ts}^*}\Big[ &-\ln\frac{\mu_W}{m_{W}}-\frac{1}{4}+\frac{1}{2(x_c-1)(x_t-1)}\,\nonumber\\
 &+\frac{x_c^3}{2(x_c-1)^2(x_c-x_t)}\ln x_c+\frac{x_t^3}{2(x_t-1)^2(x_t-x_c)}\ln x_t\, \Big]\,
\end{align}
for $b\to s\gamma$ transition, and
\begin{align}\label{C7tur}
 C_{7}^{{\rm NP}}(\mu_W) =\kappa_{tuR}^\gamma\, \frac{m_t}{\Lambda}\,\frac{V_{ud}^*}{V_{td}^*}\Big[
 &-\ln\frac{\mu_W}{m_{W}}-\frac{1}{4}+\frac{1}{2(x_u-1)(x_t-1)}\,\nonumber\\
 &+\frac{x_u^3}{2(x_u-1)^2(x_u-x_t)}\ln x_u +\frac{x_t^3}{2(x_t-1)^2(x_t-x_u)}\ln x_t\, \Big]\,
\end{align}
for $b\to d\gamma$ transition, respectively. It is noted that the NP contribution $C_7^{{\rm NP}}(\mu_W)$ is suppressed by a mass factor $m_t/\Lambda$, but enhanced by a CKM factor $V_{qD}^*/V_{tD}^*$. Since the NP contribution does not bring about any new operators, the renormalization group evolution of the Wilson coefficients from the scale $\mu_{W}$ down to $\mu_{b}$ is just the same as that in the SM.

As a final remark, we also find that the operator $\bar q_R \sigma^{\mu \nu} t_L F_{\mu \nu}$ in the effective Lagrangian Eq.~(\ref{Lagrangian}) contributes to $b \to D \gamma$ transitions only through a term $m_{D} \bar D \sigma_{\mu\nu} (1-\gamma_{5})b$. Neglecting the light quark mass $m_{D}$, as done in the SM, the effect of $\bar q_R \sigma^{\mu \nu} t_L F_{\mu \nu}$ could be therefore safely neglected, which supports the remarks on this operator made in Sec.~\ref{Sec:NP}.

\subsection{Rare $t\to q\gamma$ decays mediated by anomalous $tq\gamma$ interactions}
\label{Sec:t2qgamma}

Since $t\to b W$ is the dominant top-quark decay mode, the branching ratios of radiative $t\to q\gamma$ decays are usually defined as
\begin{equation}\label{eq:BRt2qg}
\mathcal B(t\to q\gamma) = \frac{\Gamma(t\to q \gamma)}{\Gamma(t\to b W)}\,.
\end{equation}
As the SM predictions for $\Gamma(t \to q \gamma)$ are exceedingly small~\cite{Eilam:1990zc}, we need only consider $t\to q\gamma$ decays mediated by the anomalous $tq\gamma$ interactions, which have been recently calculated at the NLO in Refs.~\cite{Zhang:2008yn,Drobnak:2010wh}. The final result for the decay widths $\Gamma(t \to q \gamma)$ reads~\cite{Zhang:2008yn,Drobnak:2010wh}
\begin{equation}
\Gamma(t\to q \gamma)=\Gamma_0(t\to q \gamma)\biggl\lbrace 1+ \frac{2\alpha_s}{9\pi} \,\left[-3\ln(\frac{\mu^2}{m_t^2})-2\pi^2+8\right]\,\biggr\rbrace,
\end{equation}
where $\Gamma_0(t \to q \gamma)= \alpha_{e}\, m_t^3\,|\kappa^{\gamma}_{\rm{tqR}}/ \Lambda|^2$, is the leading order~(LO) decay width.

The decay width of the dominant top-quark decay mode $t\to b W$ at the LO and the NLO could be found in Ref.~\cite{Li:1990qf}, and is given below
\begin{align}
\Gamma(t\to b W)=\Gamma_0(t \to b W) \biggl\lbrace  1+\frac{2\alpha_s}{3\pi}\biggl[2\left(\frac{(1-\beta_W^2)(2\beta_W^2-1)(\beta_W^2-2)}
{\beta_W^4(3-2\beta_W^2)}\right)\ln(1-\beta_W^2) \nonumber \\
-\frac{9-4\beta_W^2}{3-2\beta_W^2}\ln\beta_W^2 +2\mathrm{Li}_2(\beta_W^2) -2\mathrm{Li}_2(1-\beta_W^2)-\frac{6\beta_W^4-3\beta_W^2-8}{2\beta_W^2(3-2\beta_W^2)}-\pi^2 \biggr]\biggr\rbrace\,,
\end{align}
where $\Gamma_0(t \to bW) = \frac{G_Fm_t^3}{8\sqrt{2}\,\pi}|V_{tb}|^2\beta_W^4(3-2\beta_W^2)$, is the LO decay width and $\beta_W=(1-m_W^2/m_t^2)^{1/2}$, is the velocity of the $W$-boson in the top-quark rest frame.

\section{Numerical results and discussions}
\label{Sec:Na}

\subsection{Numerical results for the observables in $B\to V \gamma$ decays}
\label{sec:ExpandSM}

With the theoretical framework discussed above and the input parameters collected in the appendix, we first present the SM predictions for the CP-averaged branching ratios, the direct CP asymmetries, and the isospin asymmetry in $B\to K^*\gamma$ and $B\to\rho\gamma$ decays, respectively, which are listed in Table~\ref{tab:SMPredictions}. The theoretical uncertainties are obtained by varying the input parameters within their respective ranges. The experimental data are taken from the Heavy Flavor Average Group~\cite{Asner:2010qj}.

%%%%%%%%%%%%%%%%%%%%%%%%%%%%%%%%%%%%%%%%%%%%%%%%%%%%%%%%%%%%%%%%%%%
\begin{table}[t]
\begin{center}
\caption{\label{tab:SMPredictions} \small Observables in $B\to K^*\gamma$ and $B\to \rho\gamma$ decays. The branching ratios are given in unit of $10^{-6}$, while the CP and isospin asymmetries are given in unit of $10^{-2}$.}
\vspace{0.2cm}
\doublerulesep 0.8pt \tabcolsep 0.30in
\begin{tabular}{lrr}
\hline \hline
Observables & Exp. data~\cite{Asner:2010qj} & SM prediction \\
\hline
  $\mathcal B(B^+ \to K^{*+}\gamma)$ & $42.1\pm 1.8 $ & $44.60^{+12.58}_{-11.39}$ \\
  $\mathcal B(B^0 \to K^{*0}\gamma)$ & $43.3\pm 1.5 $ & $45.99^{+12.30}_{-11.22}$ \\
  $\mathcal A_{CP}(K^{*+}\gamma)$    & $  18\pm 29  $ & $-0.15^{+0.21 }_{-0.22 }$ \\
  $\mathcal A_{CP}(K^{*0}\gamma)$    & $ -16\pm 23  $ & $ 0.36^{+0.30 }_{-0.29 }$ \\
  $\Delta(K^*\gamma)$                & $ 5.2\pm 2.6 $ & $ 5.1 ^{+2.83 }_{-2.36 }$ \\
\hline
  $\mathcal B(B^+ \to \rho^{+}\gamma)$ & $0.98^{+0.25}_{-0.24}$ & $ 1.58^{+0.51}_{-0.45}$ \\
  $\mathcal B(B^0 \to \rho^{0}\gamma)$ & $0.86^{+0.15}_{-0.14}$ & $ 0.80^{+0.25}_{-0.22}$ \\
  $\mathcal A_{CP}(\rho^{+}\gamma)$    & $-11 \pm 33$           & $-9.39^{+2.55}_{-3.85}$ \\
  $\mathcal A_{CP}(\rho^{0}\gamma)$    & $\ldots$               & $-8.94^{+2.26}_{-3.08}$ \\
  $\Delta(\rho\gamma)$                 & $-46^{+17}_{-16}$      & $-8.40^{+5.41}_{-5.64}$ \\
\hline \hline
\end{tabular}
\end{center}
\end{table}
%%%%%%%%%%%%%%%%%%%%%%%%%%%%%%%%%%%%%%%%%%%%%%%%%%%%%%%%%%%%%%%%%%%

From Table~\ref{tab:SMPredictions}, we can see that, with their respective uncertainties taken into account, the SM predictions for the branching ratios are in good agreement with the experimental measurements. The predicted signs of the direct CP asymmetries in $B\to K^*\gamma$ decays are opposite to the experimental measurements, but with large uncertainties. It is noted that, within the QCDF formalism, the main theoretical uncertainty for the branching ratio comes from the tensor form factor, whereas for the direct CP asymmetry, the largest theoretical uncertainty is due to the residual renormalization-scale dependence~(for $B\to \rho\gamma$) and the first Gegenbauer moment $a_1^{\perp}$~(for $B\to K^{\ast}\gamma$). Since the isospin asymmetry in these decays is a power-suppressed effect, its calculation is less certain; however, this observable can usually provide useful information about NP parameter spaces complementary to the inclusive mode $B\to X_s\gamma$~\cite{B2Vg-NP}

As the current experimental data and the theoretical predictions for these observables still have large uncertainties, in the following numerical analyses, we shall consider them with $1\sigma$ theoretical and $2\sigma$ experimental uncertainties, respectively.

\subsection{The anomalous coupling $\kappa_{tcR}^\gamma$ in exclusive $B\to K^*\gamma$ decays}
\label{sec:B2KVg}

For $B\to K^*\gamma$ decays, the main contribution is due to the anomalous coupling $\kappa_{tcR}^\gamma$. With the notation $\kappa_{tcR}^\gamma=|\kappa_{tcR}^\gamma|\,e^{i\theta_{tcR}^\gamma}$, we get numerically
\begin{equation}\label{C7eff-tcr}
 C^{\prime\,{\rm eff}}_{7,b\to s\gamma}(\mu_b) = -0.3179 + 2.3985\,e^{i(-178.96^\circ +
 \theta_{tcR}^\gamma)}\,\frac{|\kappa_{tcR}^\gamma|}{\Lambda}\,,
\end{equation}
at the LL approximation, where the NP scale $\Lambda$ is given in unit of ${\rm TeV}$. This indicates that, for a given value $|\kappa_{tcR}^\gamma|/\Lambda$, the NP contribution is constructive to the SM one in the region $\theta_{tcR}^\gamma \simeq 0^\circ$, whereas in the regions $\theta_{tcR}^\gamma \simeq \pm 180^{\circ}$, the interference between them becomes destructive.

\begin{figure}[t]
\centering
\subfigure[$\mathcal B(B^+ \to K^{*+}\gamma)$]{\includegraphics[width=7cm]{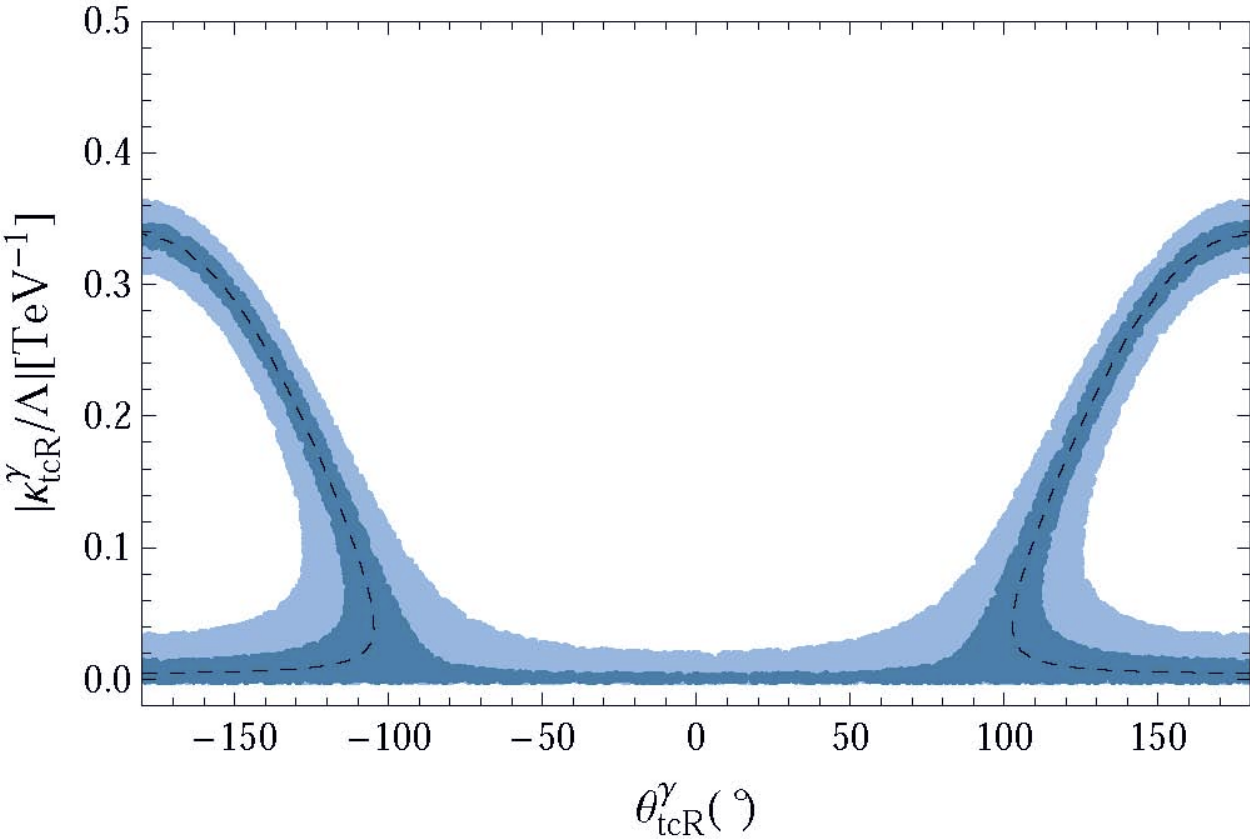} \label{fig:plotB2KVg-BRBKpg}}\hspace{0.5in}
\subfigure[$\mathcal B(B^0 \to K^{*0}\gamma)$]{\includegraphics[width=7cm]{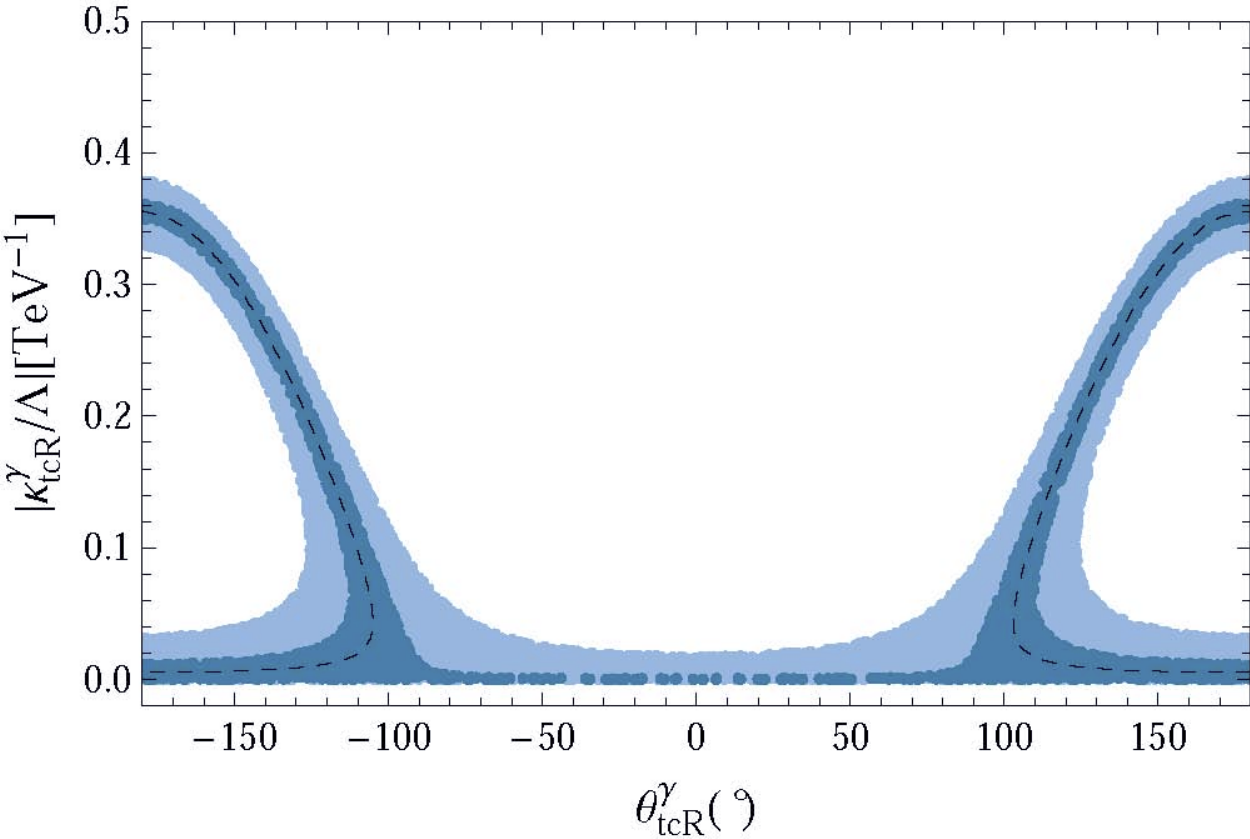}
\label{fig:plotB2KVg-BRBKzg}}\vspace{0.2in}
\subfigure[$\mathcal A_{CP}(K^{*+}\gamma)$]{\includegraphics[width=7cm]{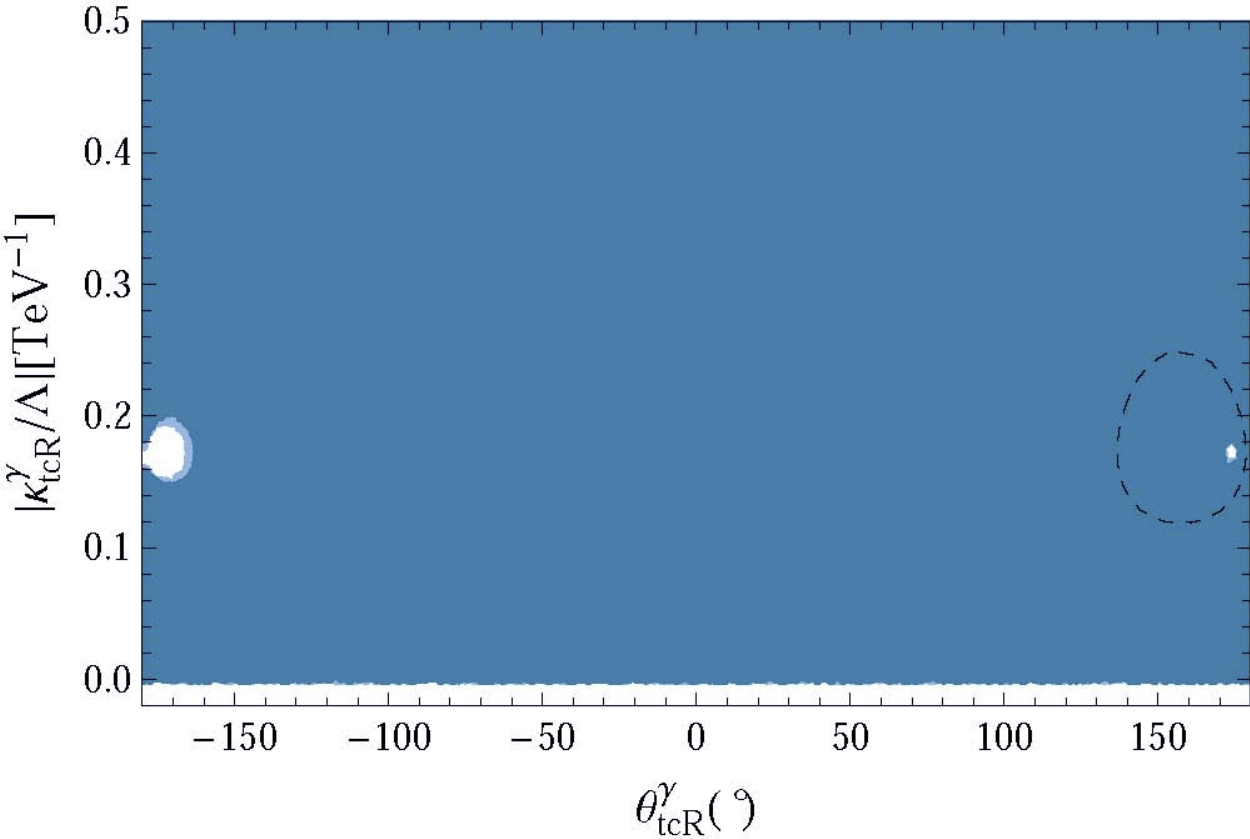} \label{fig:plotB2KVg-ACPKpg}}\hspace{0.5in}
\subfigure[$\mathcal A_{CP}(K^{*0}\gamma)$]{\includegraphics[width=7cm]{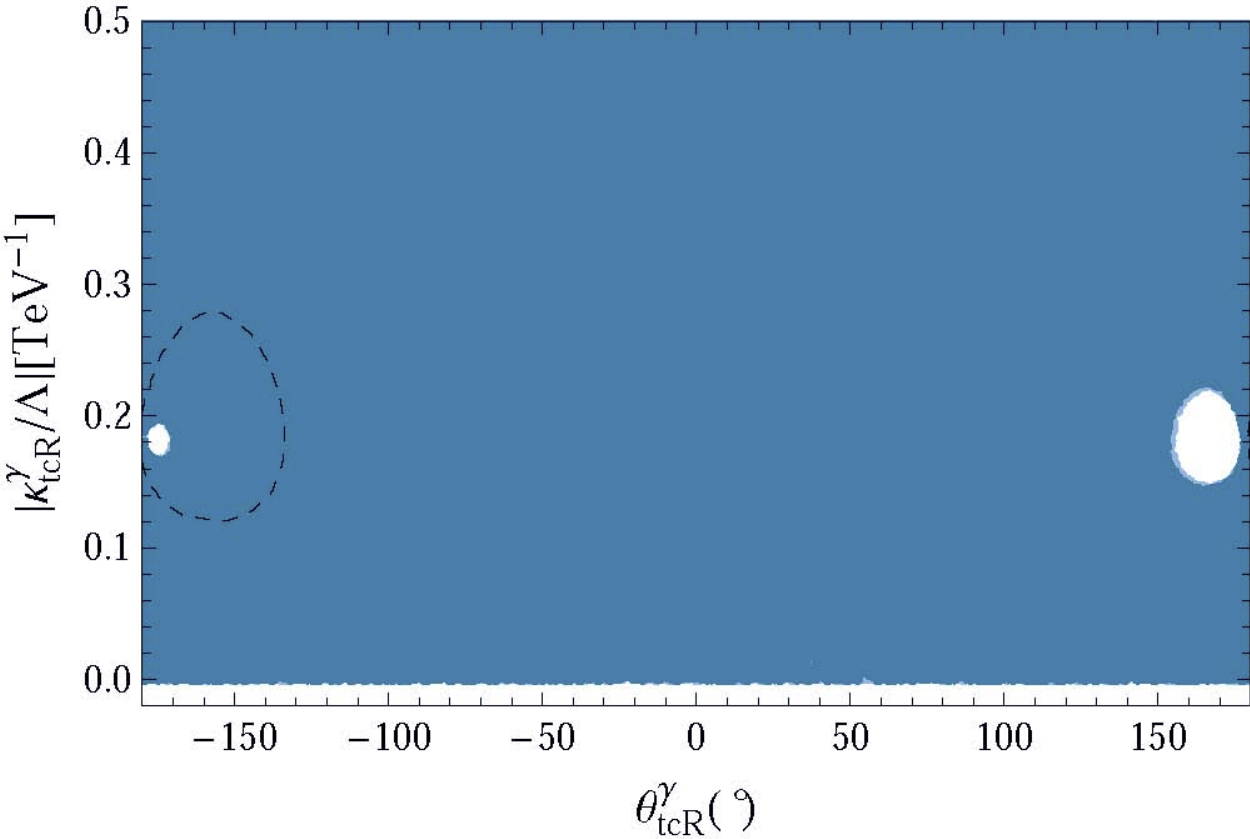}
\label{fig:plotB2KVg-ACPKzg}}\vspace{0.2in}
\subfigure[$\Delta(K^*\gamma)$]{\includegraphics[width=7cm]{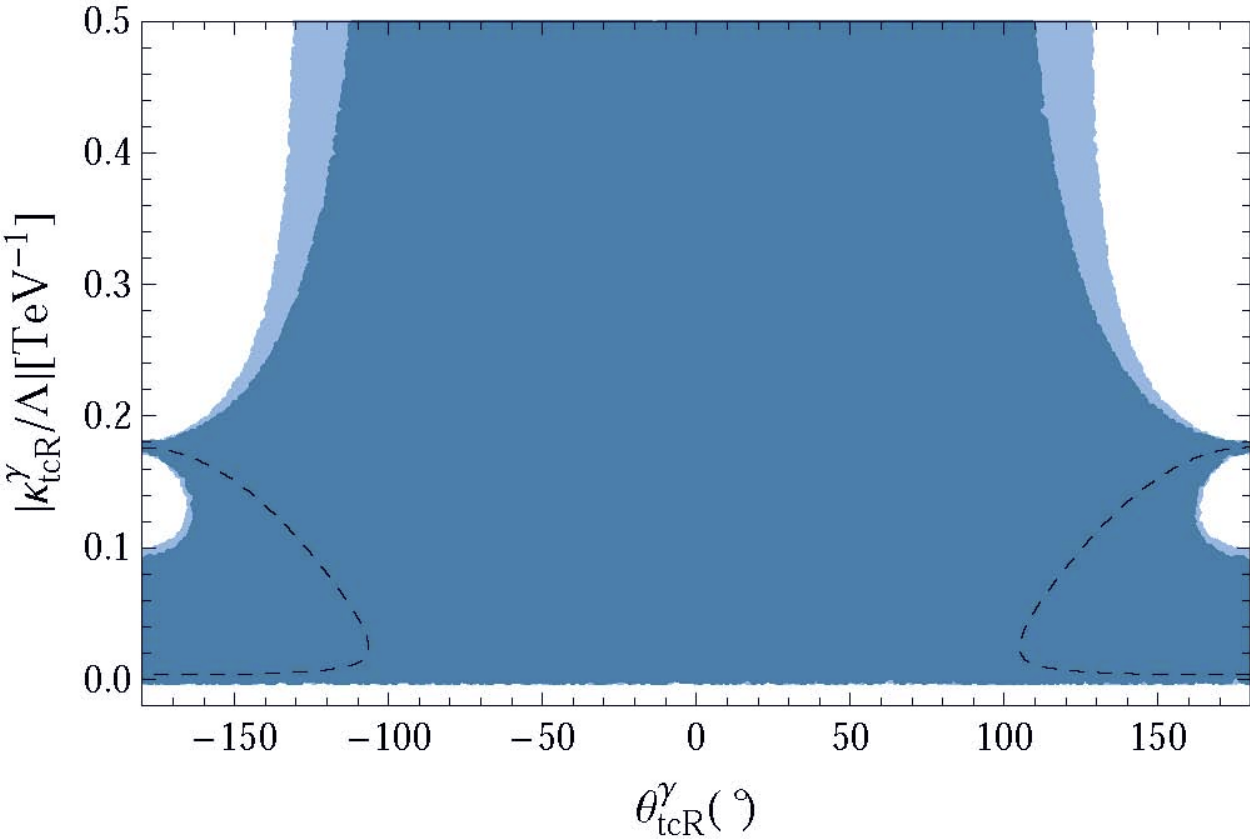}
\label{fig:plotB2KVg-DelKg}}\hspace{0.5in}
\subfigure[Combined constraints]{\includegraphics[width=7cm]{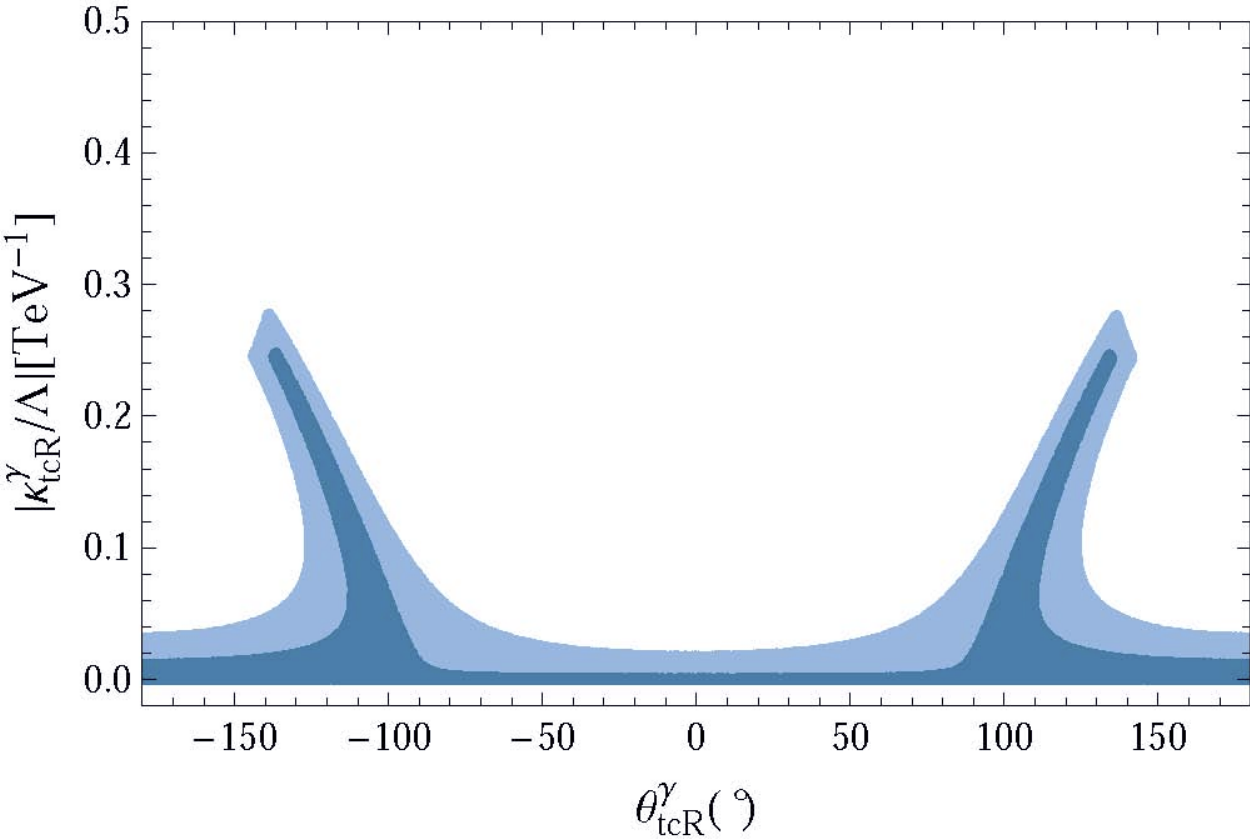}
\label{fig:plotB2KVg-combinedconstraints}}
\caption{\small The upper bounds on the anomalous coupling $|\kappa_{tcR}^{\gamma}/\Lambda|$ as a function of $\theta_{tcR}^\gamma$, constrained by the five observables in $B\to K^* \gamma$ decays. The light and the dark blue region are obtained with $1\sigma$ theoretical and $2\sigma$ experimental uncertainties, and with only $2\sigma$ experimental uncertainty, respectively. The dashed curves are obtained with central values of the input parameters.} \label{fig:plotB2KVg}
\end{figure}

The upper bounds on the anomalous coupling $|\kappa_{tcR}^{\gamma}/\Lambda|$ as a function of $\theta _{tcR}^\gamma$, constrained by the five observables in $B\to K^*\gamma$ decays, are shown in Fig.~\ref{fig:plotB2KVg}. The light and the dark blue region are obtained with $1\sigma$ theoretical and $2\sigma$ experimental uncertainties, and with only $2\sigma$ experimental uncertainty, respectively. The dashed curves are obtained with central values.

%%%%%%%%%%%%%%%%%%%%%%%%%%%%%%%%%%%%%%%%%%%%%%%%%%%%%%%%%%%%%%%%%%%
%%%%%%%%%%%%%%%%%%%%%%%%%%%%%%%%%%%%%%%%%%%%%%%%%%%%%%%%%%%%%%%%%%%

From Figs.~\ref{fig:plotB2KVg-BRBKpg} and \ref{fig:plotB2KVg-BRBKzg}, we can see that the constraints from the two branching ratios are quite similar to, but slightly looser than the one from $\mathcal B(B\to X_s\gamma)$~\cite{Yuan:2010vk}. This is due to the fact that both the exclusive and the inclusive branching ratio are, at the LO approximation, proportional to the same Wilson coefficient $|C_7|^2$. The large theoretical uncertainty is also clear by comparing the light blue region with the dashed curve. In addition, the region $\theta_{tcR}^\gamma \simeq 0$, where the NP contribution is constructive to the SM one, gives the most stringent upper bound on the magnitude $|\kappa_{tcR}^{\gamma}/\Lambda|$. There are, however, two allowed solutions in the regions $\theta_{tcR}^\gamma \simeq \pm 180^{\circ}$, the larger one corresponding to the case in which the sign of $C_7^{\rm eff}$ is flipped.

Due to the large theoretical and experimental uncertainties, the two direct CP asymmetries currently provide no valuable information about the anomalous coupling, as shown in Figs.~\ref{fig:plotB2KVg-ACPKpg} and \ref{fig:plotB2KVg-ACPKzg}. However, once measured and computed precisely, as shown by the dashed curves, these observables are also expected to provide very useful constraints.

Although being a power-suppressed effect, the isospin asymmetry, being proportional to $C_7^{-1}$ to first order, can provide complementary constraints on the anomalous top-quark coupling, as shown in Fig.~\ref{fig:plotB2KVg-DelKg}. Especially, the large sign-flipped solution allowed by the two branching ratios can be, to some extent, reduced. Since we have included  $2\sigma$ uncertainty of the experimental data, the constraint from this observable is quite loose, particularly in the region $\theta_{tcR}^\gamma \simeq 0$.

Combining all the five constraints from $B\to K^*\gamma$ decays, we present our final results in Fig.~\ref{fig:plotB2KVg-combinedconstraints}. We can see that there is actually no constraint on the phase $\theta_{tcR}$, while the upper bound on the strength $|\kappa_{tcR}^{\gamma}/\Lambda|$ is more stringent than obtained from the inclusive $B\to X_s\gamma$ decay. The implication for radiative $t\to c\gamma$ decay will be discussed later.

\subsection{The anomalous coupling $\kappa_{tuR}^\gamma$ in exclusive $B\to\rho\gamma$ decays}
\label{sec:B2rhog}

For $B\to \rho \gamma$ decays, the main contribution is due to the anomalous coupling $\kappa_{tuR}^\gamma$, and we get numerically
\begin{equation}\label{C7eff-tur}
 C^{\prime\,{\rm eff}}_{7,b\to d\gamma}(\mu_b) = -0.3179 + 11.3477\,e^{i(-21.78^\circ +
 \theta_{tuR}^\gamma)}\,\frac{|\kappa_{tuR}^\gamma|}{\Lambda}\,,
\end{equation}
at the LL approximation. In contrast to Eq.~(\ref{C7eff-tcr}), the destructive interference between the NP and the SM contribution occurs in the regions $\theta_{tuR}^\gamma \simeq 22^\circ$. The NP contribution is also accompanied with a larger coefficient, which is due to the larger CKM factor $|V_{ud}/V_{td}|$.

\begin{figure}[t]
\centering
\subfigure[$\mathcal B(B^+ \to \rho^{+}\gamma)$]{\includegraphics[width=7cm]{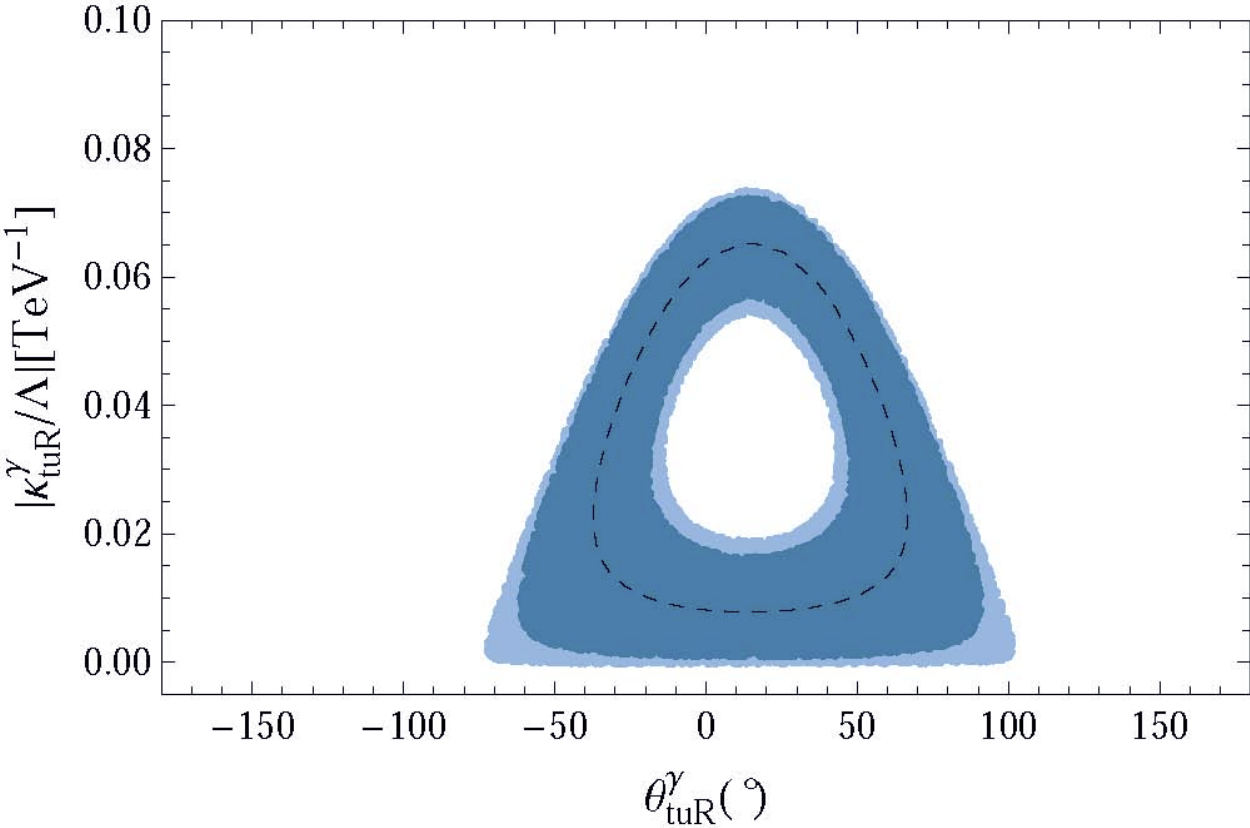}
\label{fig:plotB2Rhog-BRBRhopg}}\hspace{0.5in}
\subfigure[$\mathcal B(B^0 \to \rho^{0}\gamma)$]{\includegraphics[width=7cm]{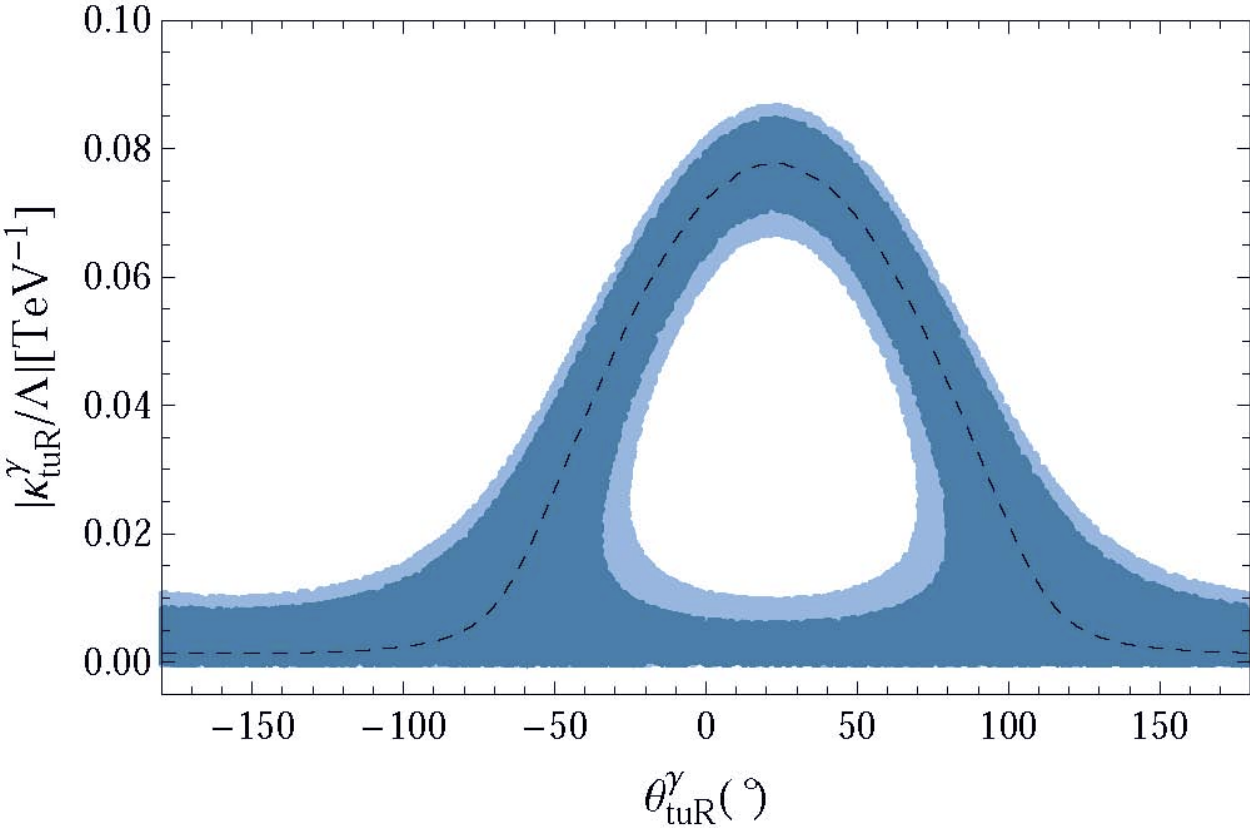}
\label{fig:plotB2Rhog-BRBRhozg}}\vspace{0.2in}
\subfigure[$\mathcal A_{CP}(\rho^{+}\gamma)$]{\includegraphics[width=7cm]{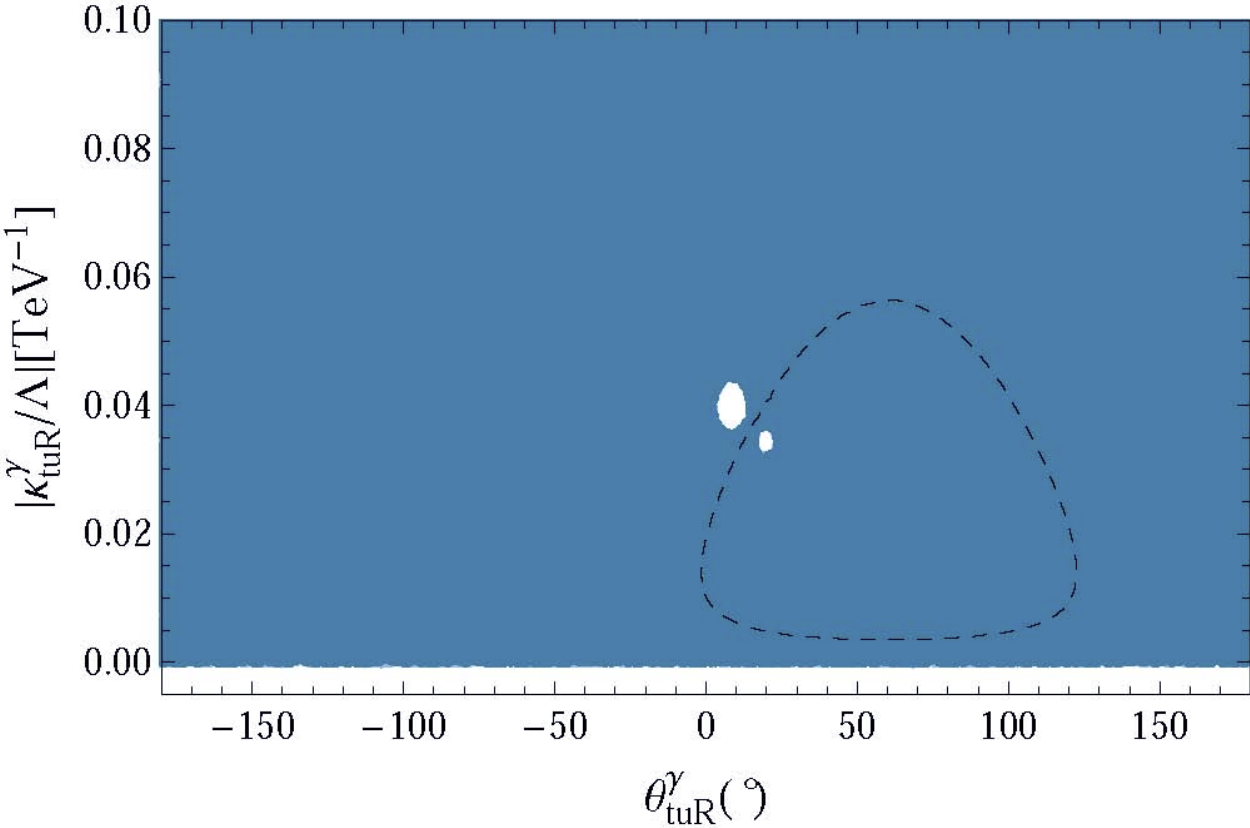}
\label{fig:plotB2Rhog-ACPRhopg}}\hspace{0.5in}
\subfigure[$\Delta(\rho\gamma)$]{\includegraphics[width=7cm]{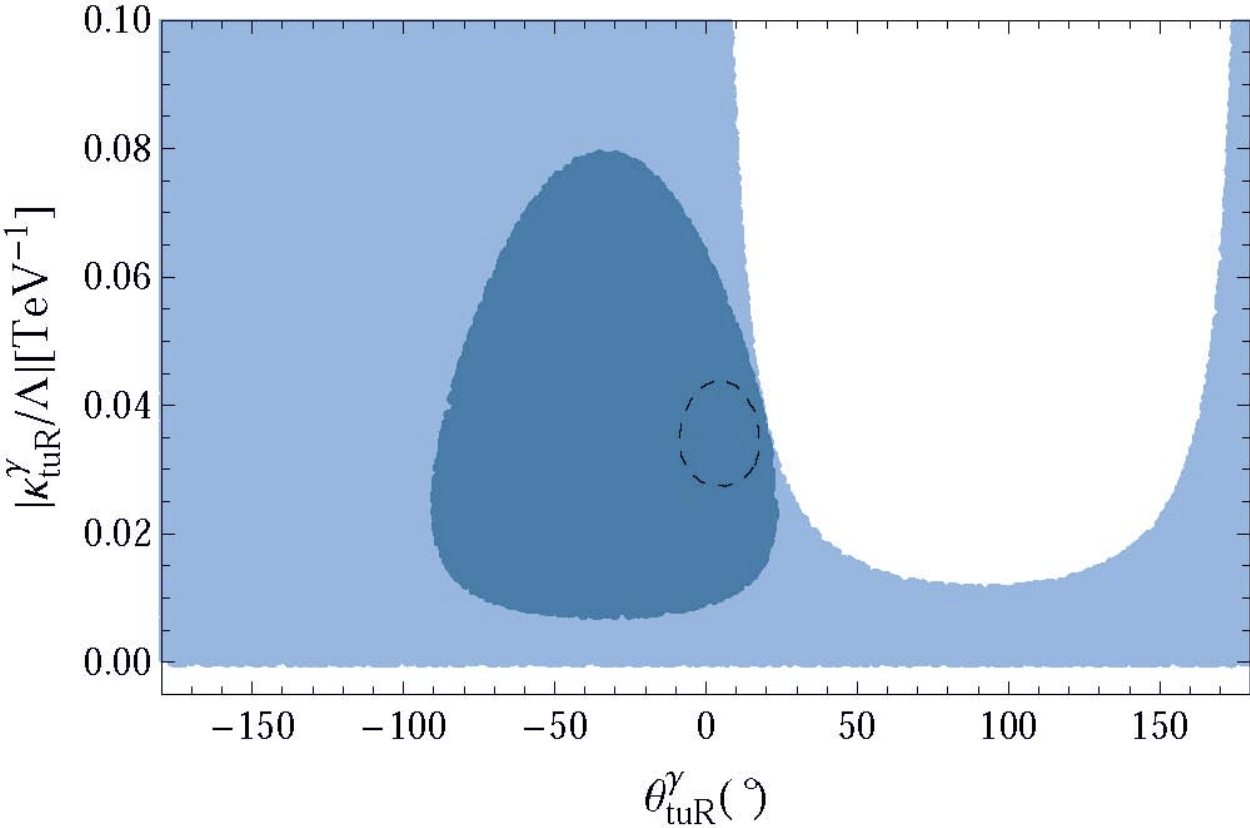}
\label{fig:plotB2Rhog-DelRhog}}\vspace{0.2in}
\subfigure[Combined constraints]{\includegraphics[width=7cm]{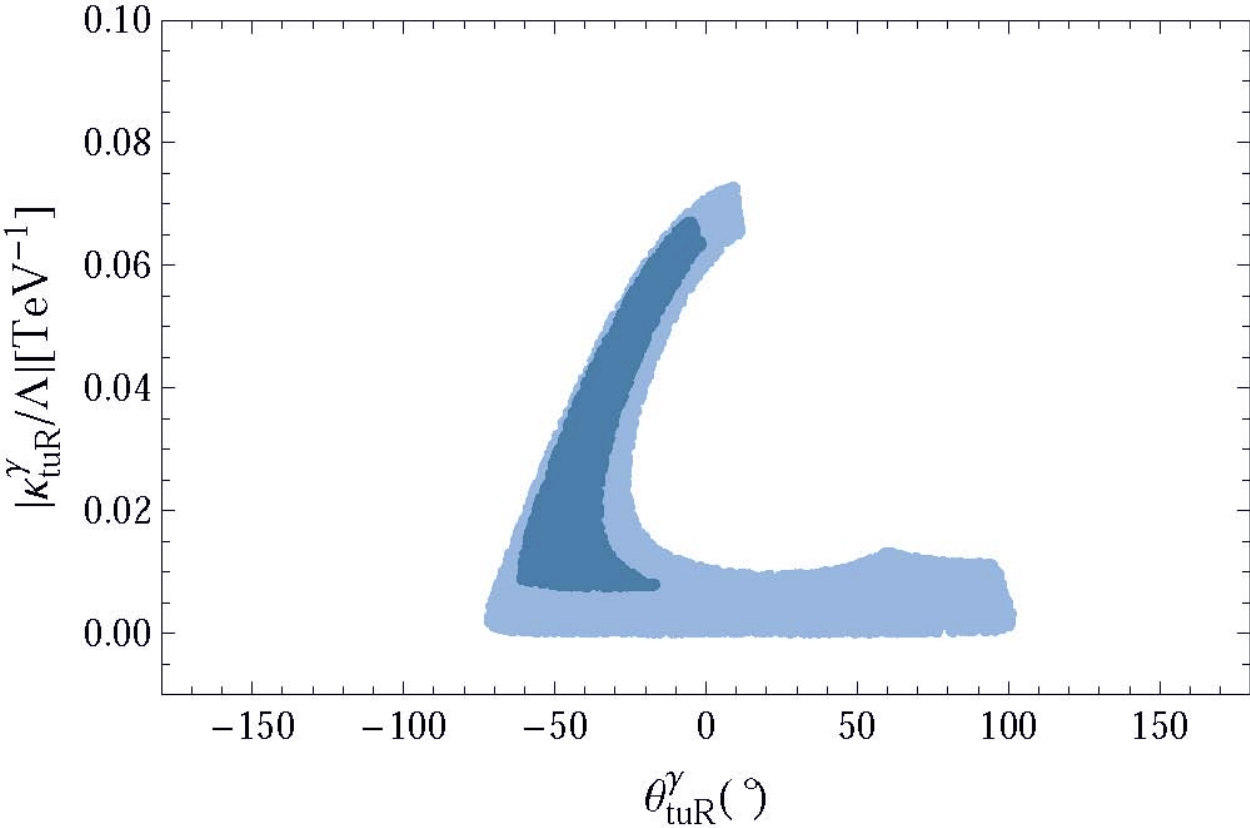}
\label{fig:plotB2Rhog-combinedconstraints}}
\caption{\small The upper bounds on anomalous coupling $|\kappa_{tuR}^{\gamma}/\Lambda|$ as a function of $\theta _{tuR}^\gamma$, constrained from the four observables in $B\to \rho \gamma$ decays. The other captions are the same as in Fig.~\ref{fig:plotB2KVg}.}
\label{fig:plotB2Rhog}
\end{figure}

As done in the case of $B\to K^* \gamma$ decays, we show in Fig.~\ref{fig:plotB2Rhog} the upper bounds on the anomalous coupling $|\kappa_{tuR}^{\gamma}/\Lambda|$ as a function of $\theta _{tuR}^\gamma$, constrained by the four observables in $B\to \rho\gamma$ decays. Due to the numerical difference between Eqs.~(\ref{C7eff-tcr}) and (\ref{C7eff-tur}), the behavior of the constraints on the NP parameters is quite different.

%%%%%%%%%%%%%%%%%%%%%%%%%%%%%%%%%%%%%%%%%%%%%%%%%%%%%%%%%%%%%%%%%%%
%%%%%%%%%%%%%%%%%%%%%%%%%%%%%%%%%%%%%%%%%%%%%%%%%%%%%%%%%%%%%%%%%%%

As shown in Fig.~\ref{fig:plotB2Rhog-BRBRhozg}, constraints from the branching ratio $\mathcal B(B^0 \to \rho^{0}\gamma)$ exhibit the destructive~(near the region $\theta_{tuR}^\gamma \simeq 22^\circ$) and constructive~(near the region $\theta_{tuR}^\gamma \simeq -158^\circ$) features between the NP and the SM contribution. The loose  bound in the destructive region corresponds to the case in which the sign of $C_7^{\rm eff}$ is flipped. On the contrary, since the SM prediction is much larger than the experimental data~(see Table~\ref{tab:SMPredictions}), the constructive region is already excluded by the branching ratio $\mathcal B(B^+ \to \rho^{+}\gamma)$, as shown in Fig.~\ref{fig:plotB2Rhog-BRBRhopg}. Thus, the phase of the anomalous coupling $\theta_{tuR}^\gamma$ is constrained into a small region by this observable.

Similar to the case in $B\to K^* \gamma$ decays, the direct CP asymmetries, with the current experimental measurements, could not give any constraints on the anomalous coupling $\kappa_{tuR}^\gamma$, which is shown in Fig.~\ref{fig:plotB2Rhog-ACPRhopg}. With respect to the isospin asymmetry $\Delta(\rho\gamma)$, as shown in Fig.~\ref{fig:plotB2Rhog-DelRhog}, it is found that the upper bounds on $|\kappa_{tuR}^{\gamma}/\Lambda|$ are more restrictive in the region $\theta_{tuR}^\gamma \in [0,180^\circ]$, while quite loose in the region $\theta_{tuR}^\gamma \in [-180^\circ,0]$. It is, however, expected to provide interesting constraints once the data becomes more precise.

The final combined constraints are shown in Fig.~\ref{fig:plotB2Rhog-combinedconstraints}. Although being excluded in the region $\theta_{tuR}^\gamma \in [0,180^\circ]$, the large sign-flipped solution is still left in the region $\theta_{tuR}^\gamma \in [-180^\circ,0]$. As already discussed, being enhanced by the large CKM factor $V_{ud}^*/V_{td}^*$, the anomalous coupling $\kappa_{tuR}^\gamma$ is expected to be severely constrained by these decay modes. However, due to the large theoretical and experimental uncertainties, the obtained upper bound on $|\kappa_{tuR}^\gamma|$ is at the same order as $|\kappa_{tcR}^\gamma|$ obtained from $B\to K^* \gamma$ decays.

\subsection{Combined upper bounds on $tq\gamma$ and implications for $\mathcal B(t\to q\gamma)$}
\label{sec:t2qgamma}

Taking $\theta_{tqR}^\gamma=0^\circ$ and $\pm 180^\circ$ as benchmarks and setting $\Lambda=1~{\rm TeV}$, we summarize our numerical constraints on the strength $|\kappa_{tqR}^{\gamma}|$ in Tables~\ref{tab:B2KVg} and \ref{tab:B2Rhog}, where S1 and S2 correspond to the two cases in which the sign of $C_7^{\rm eff}$ is not flipped and flipped, respectively. For comparisons, the constraint from $B\to X_s\gamma$~\cite{Yuan:2010vk}, and the upper bounds at $95\%$ C.L., $|\kappa_{tcR}^\gamma|<1.089$ from CDF~\cite{Abe:1997fz} and $|\kappa_{tuR}^\gamma|<0.469$ from ZEUS~\cite{Chekanov:2003yt}, are also given.

%%%%%%%%%%%%%%%%%%%%%%%%%%%%%%%%%%%%%%%%%%%%%%%%%%%%%%%%%%%%%%%%%%%
\begin{table}[t]
\begin{center}
\caption{\label{tab:B2KVg} \small Constraints on $|\kappa_{tcR}^\gamma|$ from $B\to K^* \gamma$ decays, with fixed $\theta_{tcR}^\gamma$ and $\Lambda=1~{\rm TeV}$. S1 and S2 correspond to the two cases in which the sign of $C_7^{\rm eff}$ is not flipped and flipped, respectively. For comparisons, the constraint from $B\to X_s\gamma$~\cite{Yuan:2010vk} and the upper limit from CDF~\cite{Abe:1997fz} are also listed. The corresponding  upper bounds for $\mathcal B(t\to c\gamma)$ for each case are listed in the last row.}
\vspace{0.2cm}
\doublerulesep 0.8pt \tabcolsep 0.15in
\begin{tabular}{lccc}
\hline \hline
& $\theta_{tcR}^{\gamma}=0^{\circ}$ & $\theta_{tcR}^{\gamma}=\pm 180^{\circ}$~S1 & $\theta_{tcR}^{\gamma}=\pm 180^{\circ}$~S2 \\
\hline
$\mathcal B(B^+ \to K^{*+}\gamma)$  & [0, 0.018] & [0, 0.032] & [0.311, 0.361] \\
$\mathcal B(B^0 \to K^{*0}\gamma)$  & [0, 0.017] & [0, 0.032] & [0.329, 0.379] \\
$\Delta(K^*\gamma)$                 &  $\ldots$  & [0, 0.096] & [0.175, 0.177] \\
Combined                            & [0, 0.017] & [0, 0.032] & $\varnothing$  \\ \hline
$\mathcal B(B\to X_s\gamma)$~\cite{Yuan:2010vk}
                                    & [0, 0.016] & [0, 0.019] & [0.45, 0.48]   \\
CDF bounds~\cite{Abe:1997fz}        & [0, 1.089] & [0, 1.089] & [0, 1.089]     \\ \hline
$\mathcal B(t\to c\gamma)$        & $<7.79\times 10^{-6}$ &  $<2.76\times 10^{-5}$ & $\varnothing$\\
\hline\hline
\end{tabular}
\end{center}
\end{table}
%%%%%%%%%%%%%%%%%%%%%%%%%%%%%%%%%%%%%%%%%%%%%%%%%%%%%%%%%%%%%%%%%%%

%%%%%%%%%%%%%%%%%%%%%%%%%%%%%%%%%%%%%%%%%%%%%%%%%%%%%%%%%%%%%%%%%%%
\begin{table}[t]
\begin{center}
\caption{\label{tab:B2Rhog} \small Constraints on $|\kappa_{tuR}^\gamma|$ from $B\to \rho \gamma$ decays, with fixed $\theta_{tuR}^\gamma$ and $\Lambda=1~{\rm TeV}$. For a comparison, the upper limit from ZEUS~\cite{Chekanov:2003yt} is also shown. The other captions are the same as in Table~\ref{tab:B2KVg}.}
\vspace{0.2cm}
\doublerulesep 0.8pt \tabcolsep 0.15in
\begin{tabular}{lcccc}
\hline \hline
& $\theta_{tuR}^{\gamma}=0^{\circ}$~S1 & $\theta_{tuR}^{\gamma}=0^{\circ}$~S2 & $\theta_{tuR}^{\gamma}=\pm 180^{\circ}$ \\
\hline
$\mathcal B(B^+ \to \rho^{+}\gamma)$  & [0, 0.020] & [0.051, 0.071] & $\varnothing$ \\
$\mathcal B(B^0 \to \rho^{0}\gamma)$  & [0, 0.011] & [0.060, 0.081] & [0, 0.010]    \\
Combined                              & [0, 0.011] & [0.060, 0.071] & $\varnothing$ \\ \hline
ZEUS bounds~\cite{Chekanov:2003yt}    & [0, 0.469] & [0,     0.469] & [0, 0.469]    \\ \hline
$\mathcal B(t\to u\gamma)$            &$<3.26\times 10^{-6}$ & [$9.71\times 10^{-5}$, $1.36\times 10^{-4}$]&$\varnothing$\\
\hline\hline
\end{tabular}
\end{center}
\end{table}
%%%%%%%%%%%%%%%%%%%%%%%%%%%%%%%%%%%%%%%%%%%%%%%%%%%%%%%%%%%%%%%%%%%

From Tables~\ref{tab:B2KVg} and \ref{tab:B2Rhog}, we can see that, the strengths $|\kappa_{tcR}^\gamma|$ and $|\kappa_{tuR}^\gamma|$ constrained by these exclusive decay modes are both lower than the current experimental limits. As shown in the last column of Table~\ref{tab:B2KVg}, the large sign-flipped solution allowed by the two branching ratios does not survive the constraint from the isospin asymmetry $\Delta(K^*\gamma)$. It is also noted that, for a real coupling $\kappa_{tcR}^\gamma$, the inclusive $B\to X_s \gamma$ decay provides more restrictive bounds than the exclusive $B\to K^* \gamma$ decays.

\begin{figure}[t]
\centering
 \subfigure[$\mathcal B(t\to c\gamma)$]{\includegraphics[width=7cm]{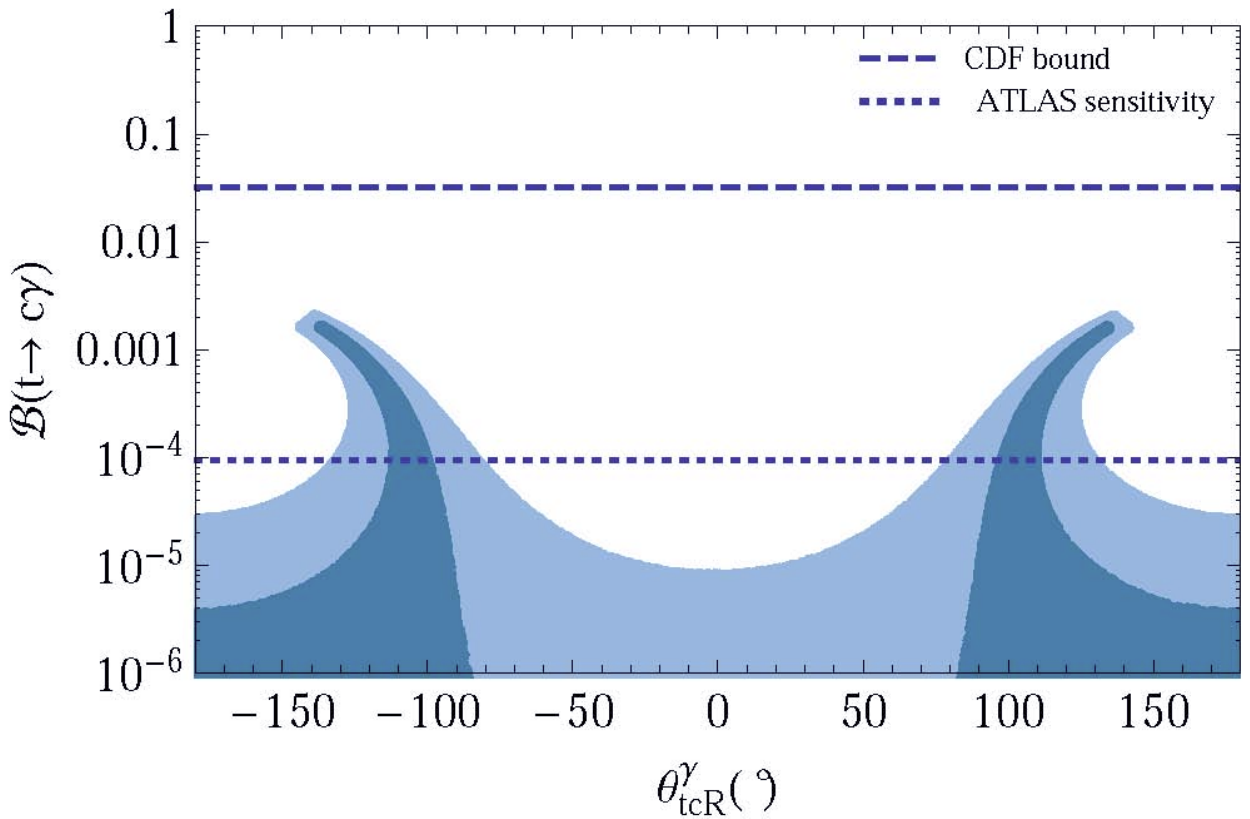}}\label{fig:t2qg-boundsK} \hspace{0.5in}
 \subfigure[$\mathcal B(t\to u\gamma)$]{\includegraphics[width=7cm]{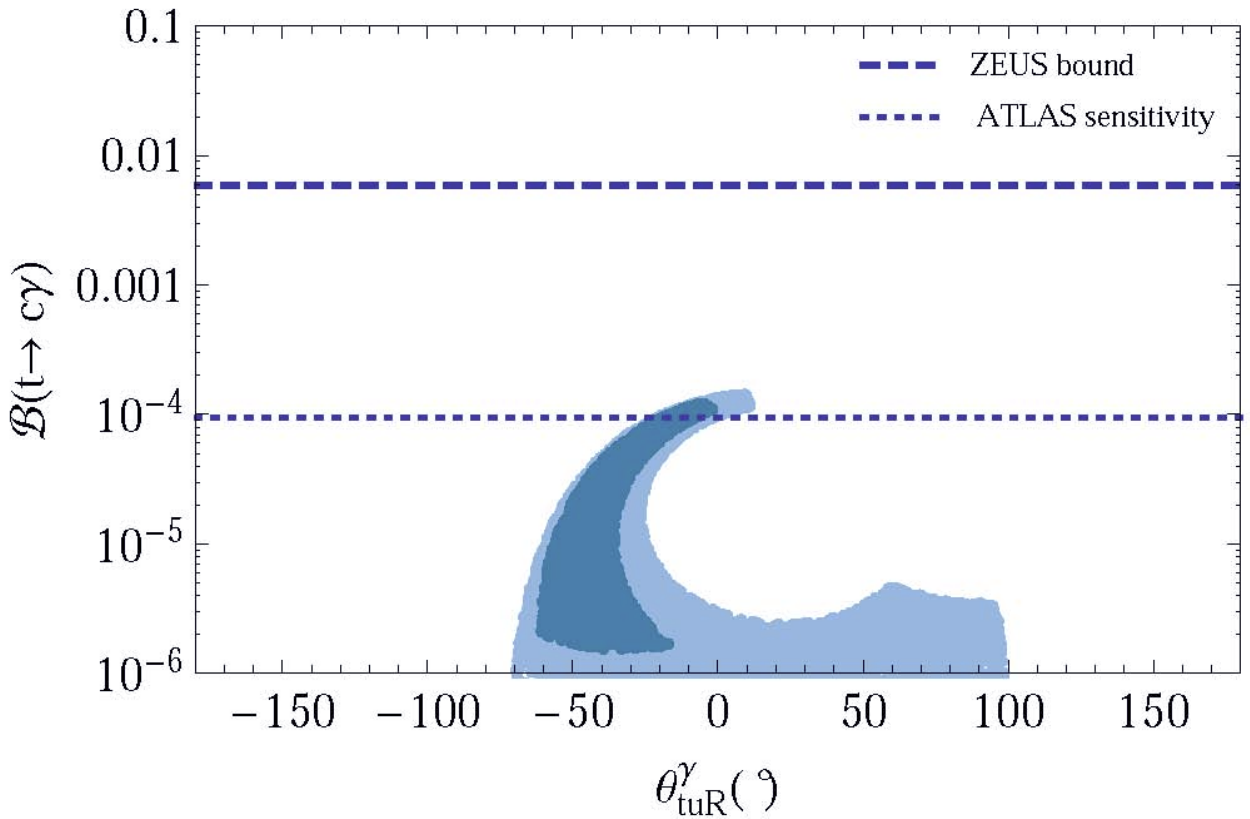}\label{fig:t2qg-boundsrho}}
\caption{\small The upper bound on $\mathcal B(t\to q\gamma)$ as a function of $\theta_{tqR}^\gamma$. Both the CDF~\cite{Abe:1997fz} and the ZEUS~\cite{Chekanov:2003yt} upper limit are given at $95\%$ C.L.. The ATLAS sensitivity~\cite{Carvalho:2007yi}, with a $5\sigma$ significance at an integrated luminosity of $10~{\rm fb}^{-1}$, is also shown.}\label{fig:t2qg}
\label{summary1}
\end{figure}

Finally, we collect in Table~\ref{tab:combined} the combined upper bounds for the anomalous couplings $tq\gamma$, and the corresponding predictions for $\mathcal B(t\to q\gamma)$. The dependence of $\mathcal B(t\to q\gamma)$ on the phase $\theta_{tqR}^{\gamma}$ is shown in Fig.~\ref{fig:t2qg}, where for comparisons, the $95\%$ C.L. upper limit $\mathcal B(t\to q\gamma)<3.2\%$ from CDF~\cite{Abe:1997fz} and $\mathcal B(t\to u\gamma)<0.59\%$ from ZEUS~\cite{Chekanov:2003yt}, as well as the ATLAS sensitivity with a $5\sigma$ significance~\cite{Carvalho:2007yi}, $\mathcal B(t\to q\gamma)<9.4\times 10^{-5}$, are also given.

%%%%%%%%%%%%%%%%%%%%%%%%%%%%%%%%%%%%%%%%%%%%%%%%%%%%%%%%%%%%%%%%%%%
\begin{table}[t]
\begin{center}
\caption{\label{tab:combined} \small Combined upper bounds for the anomalous couplings $tq\gamma$ from exclusive radiative B-meson decays, and the corresponding predictions for $\mathcal B(t\to q\gamma)$. For comparisons, we also list the upper bound from $B\to X_s\gamma$~\cite{Yuan:2010vk}, CDF~\cite{Abe:1997fz} and ZEUS~\cite{Chekanov:2003yt}, respectively.}
\vspace{0.2cm}
\doublerulesep 0.8pt \tabcolsep 0.08in
\begin{tabular}{lcc|lc}
\hline\hline
& $B\to K^* \gamma $ & $B\to X_s \gamma$~\cite{Yuan:2010vk} & & $B\to \rho \gamma$ \\
\hline
$\theta_{tcR}^{\gamma}~[^\circ]$ &  $\ldots$  &  $\ldots$ & $\theta_{tuR}^{\gamma}~[^\circ]$  & $[-71.6, 100.8]$ \\
$|\kappa_{tcR}^{\gamma}/\Lambda|~[{\rm TeV}^{-1}]$ & $<0.277$ & $<0.48$ &
$|\kappa_{tuR}^{\gamma}/\Lambda|~[{\rm TeV}^{-1}]$ & $<0.073$ \\ \hline
$\mathcal B(t\to c\gamma)$ & $<0.21\%$ & $<0.63\%$ & $\mathcal B(t\to u\gamma)$ & $<1.44\times 10^{-4}$ \\
$\mathcal B(t\to c\gamma)$~CDF~\cite{Abe:1997fz} & $<3.2\%$ & & $\mathcal B(t\to u\gamma)$~ZEUS~\cite{Chekanov:2003yt} & $<0.59\%$ \\
\hline\hline
\end{tabular}
\end{center}
\end{table}
%%%%%%%%%%%%%%%%%%%%%%%%%%%%%%%%%%%%%%%%%%%%%%%%%%%%%%%%%%%%%%%%%%%

%%%%%%%%%%%%%%%%%%%%%%%%%%%%%%%%%%%%%%%%%%%%%%%%%%%%%%%%%%%%%%%%%%%
%%%%%%%%%%%%%%%%%%%%%%%%%%%%%%%%%%%%%%%%%%%%%%%%%%%%%%%%%%%%%%%%%%%

From Table~\ref{tab:combined}, we can see that the combined upper bound on $|\kappa_{tcR}^{\gamma}|$ obtained from $B\to K^* \gamma$ decays is a bit smaller than that from the inclusive $B\to X_s \gamma$ decay. As mentioned already, this is mainly due to the constraint from the isospin asymmetry, which makes the large sign-flipped solution reduced. As shown in Fig.~\ref{fig:t2qg}, the predicted upper limits on $\mathcal B(t\to c\gamma)$ and $\mathcal B(t\to u\gamma)$ are lower than the ones from the CDF~\cite{Abe:1997fz} and the ZEUS~\cite{Chekanov:2003yt} collaboration, respectively. They are, however, of the same order as the $5\sigma$ discovery potential of ATLAS~\cite{Carvalho:2007yi}, with an integrated luminosity of $10~{\rm fb}^{-1}$.

\section{Conclusions}
\label{Sec:Conclusions}

In this paper, within the QCD factorization formalism, we have studied the effects of anomalous top-quark FCNC interactions $tq\gamma$ in exclusive radiative $B\to K^*\gamma$ and $B\to\rho\gamma$ decays. Among the two dimension-5 operators, $\bar q_R \sigma^{\mu \nu} t_L F_{\mu \nu}$ and $\bar q_L \sigma^{\mu \nu} t_R F_{\mu \nu}$, only the second one is found to give a main contribution to these decays. With the current experimental data of the branching ratios, the direct CP and the isospin asymmetries, bounds on the couplings $\kappa_{tqR}^{\gamma}$, which determines the strength of this operator, are derived. Our main conclusions are summarized as follows.

For $B\to K^*\gamma$ decays, the main contribution is due to the coupling $\kappa_{tcR}^{\gamma}$. The combined constraints from the two branching ratios and the isospin asymmetry exclude the region where the sign of $C_7^{\rm eff}$ is flipped due to the NP contribution. As a result, the upper bound on $|\kappa_{tcR}^{\gamma}/\Lambda|$ is stronger than that obtained from the inclusive $B\to X_s \gamma$ decay. The corresponding upper limit, $\mathcal B(t\to c\gamma)<0.21\%$, is lower than the CDF result. However,
for real $\kappa_{tcR}^{\gamma}$, the bound is rather restrictive, and the corresponding upper limit on $\mathcal B(t\to c\gamma)$ is generally of the same order as the $5\sigma$ discovery potential of ATLAS with an integrated luminosity of $10~{\rm fb}^{-1}$.

For $B\to\rho\gamma$ decays, the main contribution is, on the other hand, due to the coupling $\kappa_{tuR}^{\gamma}$ and enhanced by a large CKM factor $|V_{ud}/V_{td}|$. Due to the large theoretical and experimental uncertainties, constraints on the anomalous coupling are not as strong as expected. However, with the current data and our SM prediction of $\mathcal B(B^+\to\rho^+\gamma)$, between which  there is a large difference, the phase of the anomalous coupling $\theta_{tuR}^\gamma$ gets, to some extent, constrained. The obtained upper limit, $\mathcal B(t\to u\gamma)<1.44\times 10^{-4}$, is stronger than the ZEUS result. For most of the constrained parameter space, the upper limit on $\mathcal B(t\to u\gamma)$ is about the same order as the $5\sigma$ discovery potential of ATLAS with an integrated luminosity of $10~{\rm fb}^{-1}$.

As a result, there is an interesting interplay between exclusive radiative B-meson and rare $t\to q\gamma$ decays. With refined measurements from the LHCb and the future super-B factories, we can get close correlation between the $b\to s\gamma$ decays  and the rare $t\to q\gamma$ decays, which will be studied directly at the LHC CMS and ATLAS.

\section*{Acknowledgements}

The work was supported in part by the National Natural Science Foundation under contract Nos.~11075059, 11005032 and 11047165. X.~Q. Li was also supported in part by MEC (Spain) under Grant FPA2007-60323 and by the Spanish Consolider Ingenio 2010 Programme CPAN (CSD2007-00042).

\begin{appendix}

\section*{Appendix: Theoretical input parameters}
\label{app:input}

In this appendix, we collect all the relevant input parameters, when calculating the observables in $B \to V\gamma$ decays and the branching ratios of $t \to q\gamma$ decays mediated by the anomalous $tq\gamma$ interactions.

\subsubsection*{The basic SM parameters}

First, we need some basic SM parameters, which are, if not stated otherwise, taken from the Particle Data Group~\cite{Nakamura:2010zzi}
\begin{align}
& \alpha_s(m_Z)=0.1187\pm 0.0007, \, \alpha=1/137.036, \, G_F=1.16637\times 10^{-5}~{\rm
GeV}^{-2}, \nonumber \\
& \sin^2\theta_W=0.23146,\, m_W=80.399~{\rm GeV}, \, m_Z=91.1876~{\rm GeV}, \, m_t=173.3\pm 1.1~{\rm GeV}~\cite{:1900yx}, \nonumber \\
& m_{B^+}=5279.17~{\rm MeV}, \, m_{B^0}=5279.50~{\rm MeV}, \, m_{K^{\ast +}}=891.66~{\rm MeV}, \, m_{K^{\ast 0}}=895.94~{\rm MeV}, \nonumber \\
& m_{\rho^{+}}=m_{\rho^{0}}=775.49~{\rm MeV}, \, \tau_{B^+}=1.638~{\rm ps}, \, \tau_{B^0}=1.525~{\rm ps},
\end{align}
where $m_t$ is the top-quark pole mass. We use two-loop running for $\alpha_s$ throughout this paper.

\subsubsection*{The CKM matrix elements}

For the CKM matrix elements, we adopt the Wolfenstein parametrization~\cite{Wolfenstein:1983yz} and choose the four
parameters $A$, $\lambda$, $\rho$ and $\eta$ as fitted by the CKMfitter group~\cite{Charles:2004jd}
\begin{equation}
 A=0.812^{+0.013}_{-0.027}\,, \quad \lambda=0.22543^{+0.00077}_{-0.00077}\,, \quad
 \overline{\rho}=0.144^{+0.025}_{-0.025}\,, \quad \overline{\eta}=0.342^{+0.016}_{-0.015}\,,
\end{equation}
with $\overline{\rho}=\rho\,(1-\frac{\lambda^2}{2})$ and $\bar{\eta}=\eta\,(1-\frac{\lambda^2}{2})$.

\subsubsection*{Quark masses and meson parameters}

For the running quark masses in the ${\rm \overline{MS}}$ scheme, we take~\cite{HPQCD_quarkmass}
\begin{equation}\label{MSbar-mass}
\overline{m}_{b}(\overline{m}_{b})=4.164\pm 0.023~{\rm GeV}, \quad
\overline{m}_{c}(\overline{m}_{c})=1.273\pm 0.006~{\rm GeV}.
\end{equation}
To get the corresponding pole and running quark masses at different scales, we use the NLO ${\rm \overline{MS}}$-on-shell conversion and running formulae collected, for example, in Ref.~\cite{Chetyrkin:2000yt}.

It should be noted that the quantity ${\cal C}_7^{(i)}$, defined by Eq.~(\ref{CalC7}), depends on the $b$-quark mass renormalization scheme. Following Ref.~\cite{Beneke:2001at}, in this paper we use the potential-subtracted~(PS) scheme~\cite{Beneke:1998rk}. Converting $\overline{m}_{b}(\overline{m}_{b}) \simeq 4.164~{\rm GeV}$ to the PS mass $m_{\rm PS}(2~{\rm GeV})$, we get numerically $m_{\rm PS}(2~{\rm GeV}) \simeq 4.45~{\rm GeV}$. To give an estimate of renormalization scale uncertainty, the two renormalization scales $\mu_b$ and $\mu_{hc}$ are chosen, respectively, as
\begin{equation}
 \mu_b=4.45^{+4.45}_{-2.22}~{\rm GeV}, \quad \mu_{hc}=1.5\pm 0.6~{\rm GeV}.
\end{equation}

When discussing exclusive $B\to V \gamma$ decays, we need some parameters related to the involved mesons, such as the transition form factors, the decay constants, as well as the Gegenbauer moments of meson LCDAs. Most of these parameters are not directly known from experiment, and have to be determined by some nonperturbative methods like the QCD sum rule and lattice QCD. A summary of these parameters is listed below
\begin{align}
& T_1^{B\to K^*}(0)=0.31\pm 0.04~\cite{Ball:2006eu}, \quad
  T_1^{B\to\rho}(0)=0.27\pm 0.04~\cite{Ball:2006eu}, \quad
  f_{B_q}=192.8\pm 9.9~{\rm MeV}~\cite{Laiho:2009eu}, \nonumber \\
& \lambda_{B,+}(1.5~{\rm GeV})=485\pm 115~{\rm MeV}~\cite{Beneke:2001at}, \quad
  f_{K^*}=220\pm 5~{\rm MeV}~\cite{Ball:2006eu}, \quad
  f_{\rho}=216\pm 3~{\rm MeV}~\cite{Ball:2006eu}, \nonumber \\
& f_{K^*}^{\perp}(1~{\rm GeV})=185\pm 10~{\rm MeV}~\cite{Ball:2006eu}, \quad
  f_{\rho}^{\perp}(1~{\rm GeV})=165\pm 9~{\rm MeV}~\cite{Ball:2006eu}, \nonumber \\
& a_1(\bar K^{\ast})_{\perp,\parallel}(1~{\rm GeV})=-0.04\pm 0.03~\cite{Ball:2006eu}, \quad
  a_2(\bar K^{\ast})_{\perp,\parallel}(1~{\rm GeV})=0.15\pm 0.10~\cite{Ball:2006eu}, \nonumber \\
& a_2(\rho)_{\perp,\parallel}(1~{\rm GeV})=0.15\pm 0.07~\cite{Ball:2006eu},
\end{align}
where the tensor form factors $T_1$ evaluated at $q^2=0$, the decay constants and Gegenbauer moments of the light mesons are obtained from QCD light-cone sum rules~(LCSRs)~\cite{Ball:2006eu}. The quantity $\lambda_{B,+}$ is defined as the first inverse moment of the B-meson LCDA~\cite{Beneke:2001at}, and $f_{B_q}$ is the B-meson decay constant.

\end{appendix}


\begin{thebibliography}{100}

%\cite{Glashow:1970gm}
\bibitem{Glashow:1970gm}
  S.~L.~Glashow, J.~Iliopoulos, L.~Maiani,
  %``Weak Interactions with Lepton-Hadron Symmetry,''
  Phys.\ Rev.\  {\bf D2}, 1285-1292 (1970).

%\cite{Eilam:1990zc}
\bibitem{Eilam:1990zc}
  G.~Eilam, J.~L.~Hewett, A.~Soni,
  %``Rare decays of the top quark in the standard and two Higgs doublet models,''
  Phys.\ Rev.\  {\bf D44}, 1473-1484 (1991),  Erratum-ibid D{\bf 59}, 039901 (1999);
%\cite{DiazCruz:1989ub}
%\bibitem{DiazCruz:1989ub}
  J.~L.~Diaz-Cruz, R.~Martinez, M.~A.~Perez, A.~Rosado,
  %``FLAVOR CHANGING RADIATIVE DECAY OF THF t QUARK,''
  Phys.\ Rev.\  {\bf D41}, 891-894 (1990).

%\cite{Beneke:2000hk}
\bibitem{Beneke:2000hk}
  M.~Beneke, I.~Efthymiopoulos, M.~L.~Mangano, J.~Womersley, A.~Ahmadov, G.~Azuelos, U.~Baur, A.~Belyaev {\it et al.},
  %``Top quark physics,''
  [hep-ph/0003033];
%\cite{Bernreuther:2008ju}
%\bibitem{Bernreuther:2008ju}
  W.~Bernreuther,
  %``Top quark physics at the LHC,''
  J.\ Phys.\ G {\bf G35}, 083001 (2008).
  [arXiv:0805.1333 [hep-ph]]; and references therein.

%\cite{Abe:1997fz}
\bibitem{Abe:1997fz}
  F.~Abe {\it et al.} [ CDF Collaboration ],
  %``Search for flavor-changing neutral current decays of the top quark in $p \bar{p}$ collisions at $\sqrt{s} = 1.8$
  %TeV,''
  Phys.\ Rev.\ Lett.\  {\bf 80}, 2525-2530 (1998).

%\cite{Chekanov:2003yt}
\bibitem{Chekanov:2003yt}
  S.~Chekanov {\it et al.} [ ZEUS Collaboration ],
  %``Search for single top production in ep collisions at HERA,''
  Phys.\ Lett.\  {\bf B559}, 153-170 (2003).
  [hep-ex/0302010].

%\cite{Carvalho:2007yi}
\bibitem{Carvalho:2007yi}
  J.~Carvalho {\it et al.} [ ATLAS Collaboration ],
  %``Study of ATLAS sensitivity to FCNC top decays,''
  Eur.\ Phys.\ J.\  {\bf C52}, 999-1019 (2007).
  [arXiv:0712.1127 [hep-ex]];
%\cite{Veloso:2008zza}
%\bibitem{Veloso:2008zza}
  F.~M.~A.~Veloso,
  %``Study of ATLAS sensitivity to FCNC top quark decays,''
  CERN-THESIS-2008-106.

%\cite{Benucci:2008zz}
\bibitem{Benucci:2008zz}
  L.~Benucci, A.~Kyriakis,
  %``CMS sensitivity to top flavour changing neutral currents,''
  Nucl.\ Phys.\ Proc.\ Suppl.\  {\bf 177-178}, 258-260 (2008).

%\cite{Nakamura:2010zzi}
\bibitem{Nakamura:2010zzi}
  KNakamura {\it et al.} [ Particle Data Group Collaboration ],
  %``Review of particle physics,''
  J.\ Phys.\ G {\bf G37}, 075021 (2010).

\bibitem{FCNC-top-old}
%\cite{Han:1995pk}
%\bibitem{Han:1995pk}
  T.~Han, R.~D.~Peccei, X.~Zhang,
  %``Top quark decay via flavor changing neutral currents at hadron colliders,''
  Nucl.\ Phys.\  {\bf B454}, 527-540 (1995).
  [hep-ph/9506461];
%\cite{Han:1996ep}
%\bibitem{Han:1996ep}
  T.~Han, K.~Whisnant, B.~L.~Young, X.~Zhang,
  %``Top quark decay via the anomalous coupling $\bar{t} c \gamma$ at hadron colliders,''
  Phys.\ Rev.\  {\bf D55}, 7241-7248 (1997).
  [hep-ph/9603247];
%\cite{Han:1996ce}
%\bibitem{Han:1996ce}
  %T.~Han, K.~Whisnant, B.~L.~Young, X.~Zhang,
  %``Searching for $t \to c$ g at the Fermilab Tevatron,''
  Phys.\ Lett.\  {\bf B385}, 311-316 (1996).
  [hep-ph/9606231];
%\cite{Larios:1999au}
%\bibitem{Larios:1999au}
  F.~Larios, M.~A.~Perez, C.~P.~Yuan,
  %``Analysis of $t b W$ and $t t Z$ couplings from CLEO and LEP / SLC data,''
  Phys.\ Lett.\  {\bf B457}, 334-340 (1999).
  [hep-ph/9903394];
%\cite{Burdman:1999fw}
%\bibitem{Burdman:1999fw}
  G.~Burdman, M.~C.~Gonzalez-Garcia, S.~F.~Novaes,
  %``Anomalous couplings of the third generation in rare B decays,''
  Phys.\ Rev.\  {\bf D61}, 114016 (2000).
  [hep-ph/9906329].

%\cite{Lee:2008xr}
\bibitem{Lee:2008xr}
  J.~P.~Lee, K.~Y.~Lee,
  %``Implications of the anomalous top quark couplings in $B_s$ - $\bar{B}_s$ mixing, $B \to X_{s} \gamma$ and top quark
  %decays,''
  Phys.\ Rev.\  {\bf D78}, 056004 (2008).
  [arXiv:0806.1389 [hep-ph]];
%\cite{Lee:2006qv}
%\bibitem{Lee:2006qv}
  K.~Y.~Lee,
  %``CP violation in B ---> rho gamma decay with anomalous right-handed top quark couplings,''
  Phys.\ Lett.\  {\bf B632}, 99-104 (2006).

%\cite{Fox:2007in}
\bibitem{Fox:2007in}
  P.~J.~Fox, Z.~Ligeti, M.~Papucci, G.~Perez, M.~D.~Schwartz,
  %``Deciphering top flavor violation at the LHC with $B$ factories,''
  Phys.\ Rev.\  {\bf D78}, 054008 (2008).
  [arXiv:0704.1482 [hep-ph]].

%\cite{Grzadkowski:2008mf}
\bibitem{Grzadkowski:2008mf}
  B.~Grzadkowski, M.~Misiak,
  %``Anomalous Wtb coupling effects in the weak radiative B-meson decay,''
  Phys.\ Rev.\  {\bf D78}, 077501 (2008).
  [arXiv:0802.1413 [hep-ph]].

%\cite{Yuan:2010vk}
\bibitem{Yuan:2010vk}
  X.~Yuan, Y.~Hao, Y.~Yang,
  %``$B -> X_s\gamma$ constraints on the top quark anomalous $t-> c \gamma$ coupling,''
  Phys.\ Rev.\  {\bf D83}, 013004 (2011).
  [arXiv:1010.1912 [hep-ph]].

%\cite{Asner:2010qj}
\bibitem{Asner:2010qj}
  D.~Asner {\it et al.} [ Heavy Flavor Averaging Group Collaboration ],
  %``Averages of b-hadron, c-hadron, and $\tau-lepton Properties,''
  [arXiv:1010.1589 [hep-ex]], and online update at http://www.slac.stanford.edu/xorg/hfag.

\bibitem{B2Vgreview}
 For a recent review, see:
%\cite{Hurth:2010tk}
%\bibitem{Hurth:2010tk}
  T.~Hurth, M.~Nakao,
  %``Radiative and Electroweak Penguin Decays of B Mesons,''
  Ann.\ Rev.\ Nucl.\ Part.\ Sci.\  {\bf 60}, 645-677 (2010).
  [arXiv:1005.1224 [hep-ph]].

\bibitem{B2Vg-PQCD}
%\cite{Keum:2004is}
%\bibitem{Keum:2004is}
  Y.~Y.~Keum, M.~Matsumori, A.~I.~Sanda,
  %``CP asymmetry, branching ratios and isospin breaking effects of B ---> K* gamma with perturbative QCD approach,''
  Phys.\ Rev.\  {\bf D72}, 014013 (2005).
  [hep-ph/0406055];
%\cite{Lu:2005yz}
%\bibitem{Lu:2005yz}
  C.~-D.~Lu, M.~Matsumori, A.~I.~Sanda, M.~-Z.~Yang,
  %``CP asymmetry, branching ratios and isospin breaking effects in B ---> rho gamma and B ---> omega gamma decays with
  %the pQCD approach,''
  Phys.\ Rev.\  {\bf D72}, 094005 (2005).
  [hep-ph/0508300];
%\cite{Matsumori:2005ax}
%\bibitem{Matsumori:2005ax}
  M.~Matsumori, A.~I.~Sanda,
  %``The Mixing-induced CP asymmetry in B ---> K* gamma decays with perturbative QCD approach,''
  Phys.\ Rev.\  {\bf D73}, 114022 (2006).
  [hep-ph/0512175];
%\cite{Wang:2007an}
%\bibitem{Wang:2007an}
  W.~Wang, R.~-H.~Li, C.~-D.~Lu,
  %``Radiative charmless B(s) ---> V gamma and B(s) ---> A gamma decays in pQCD approach,''
  [arXiv:0711.0432 [hep-ph]].

\bibitem{B2Vg-SCET}
%\cite{Ali:2007sj}
%\bibitem{Ali:2007sj}
  A.~Ali, B.~D.~Pecjak, C.~Greub,
  %``$B \to$ V $\gamma$ Decays at NNLO in SCET,''
  Eur.\ Phys.\ J.\  {\bf C55}, 577-595 (2008).
  [arXiv:0709.4422 [hep-ph]];
%\cite{Kim:2008rz}
%\bibitem{Kim:2008rz}
  C.~Kim, A.~K.~Leibovich, T.~Mehen,
  %``Nonperturbative Charming Penguin Contributions to Isospin Asymmetries in Radiative B decays,''
  Phys.\ Rev.\  {\bf D78}, 054024 (2008).
  [arXiv:0805.1735 [hep-ph]];
%\cite{Becher:2005fg}
%\bibitem{Becher:2005fg}
  T.~Becher, R.~J.~Hill, M.~Neubert,
  %``Factorization in B ---> V gamma decays,''
  Phys.\ Rev.\  {\bf D72}, 094017 (2005).
  [hep-ph/0503263];
%\cite{Chay:2003kb}
%\bibitem{Chay:2003kb}
  J.~-g.~Chay, C.~Kim,
  %``Rare radiative exclusive B decays in soft collinear effective theory,''
  Phys.\ Rev.\  {\bf D68}, 034013 (2003).
  [hep-ph/0305033].

%\cite{Ball:2006eu}
\bibitem{Ball:2006eu}
  P.~Ball, G.~W.~Jones, R.~Zwicky,
  %``B ---> V gamma beyond QCD factorisation,''
  Phys.\ Rev.\  {\bf D75}, 054004 (2007).
  [hep-ph/0612081].

\bibitem{B2Vg-LCSR}
%\cite{Ball:2006nr}
%\bibitem{Ball:2006nr}
  P.~Ball, R.~Zwicky,
  %``$|V_{td} / V_{ts}|$ from $B \to V \gamma$,''
  JHEP {\bf 0604}, 046 (2006).
  [hep-ph/0603232];
%\cite{Ball:2006cva}
%\bibitem{Ball:2006cva}
  %P.~Ball, R.~Zwicky,
  %``Time-dependent CP Asymmetry in B ---> K* gamma as a (Quasi) Null Test of the Standard Model,''
  Phys.\ Lett.\  {\bf B642}, 478-486 (2006).
  [hep-ph/0609037];
%\cite{Muheim:2008vu}
%\bibitem{Muheim:2008vu}
  F.~Muheim, Y.~Xie, R.~Zwicky,
  %``Exploiting the width difference in $B_s \to \phi \gamma$,''
  Phys.\ Lett.\  {\bf B664}, 174-179 (2008).
  [arXiv:0802.0876 [hep-ph]];
%\cite{Khodjamirian:2010vf}
%\bibitem{Khodjamirian:2010vf}
  A.~Khodjamirian, T.~.Mannel, A.~A.~Pivovarov, Y.~-M.~Wang,
  %``Charm-loop effect in $B \to K^{(*)} \ell^{+} \ell^{-}$ and $B\to K^*\gamma$,''
  JHEP {\bf 1009}, 089 (2010).
  [arXiv:1006.4945 [hep-ph]].

%\cite{Beneke:2001at}
\bibitem{Beneke:2001at}
  M.~Beneke, T.~Feldmann, D.~Seidel,
  %``Systematic approach to exclusive B ---> V l+ l-, V gamma decays,''
  Nucl.\ Phys.\  {\bf B612}, 25-58 (2001).
  [hep-ph/0106067];
%\cite{Beneke:2004dp}
%\bibitem{Beneke:2004dp}
  %M.~Beneke, T.~.Feldmann, D.~Seidel,
  %``Exclusive radiative and electroweak b ---> d and b ---> s penguin decays at NLO,''
  Eur.\ Phys.\ J.\  {\bf C41}, 173-188 (2005).
  [hep-ph/0412400].

%\cite{Bosch:2001gv}
\bibitem{Bosch:2001gv}
  S.~W.~Bosch, G.~Buchalla,
  %``The Radiative decays B ---> V gamma at next-to-leading order in QCD,''
  Nucl.\ Phys.\  {\bf B621}, 459-478 (2002).
  [hep-ph/0106081];
%\cite{Bosch:2002bw}
%\bibitem{Bosch:2002bw}
  S.~W.~Bosch,
  %``Exclusive radiative decays of B mesons in QCD factorization,''
  [hep-ph/0208203].

%\cite{Ali:2001ez}
\bibitem{Ali:2001ez}
  A.~Ali, A.~Y.~Parkhomenko,
  %``Branching ratios for B ---> K* gamma and B ---> rho gamma decays in next-to-leading order in the large energy
  %effective theory,''
  Eur.\ Phys.\ J.\  {\bf C23 } (2002)  89-112.
  [hep-ph/0105302];
%\cite{Ali:2004hn}
%\bibitem{Ali:2004hn}
  A.~Ali, E.~Lunghi, A.~Y.~.Parkhomenko,
  %``Implication of the B ---> (rho, omega) gamma branching ratios for the CKM phenomenology,''
  Phys.\ Lett.\  {\bf B595}, 323-338 (2004).
  [hep-ph/0405075].

%\cite{Kagan:2001zk}
\bibitem{Kagan:2001zk}
  A.~L.~Kagan, M.~Neubert,
  %``Isospin breaking in B ---> K* gamma decays,''
  Phys.\ Lett.\  {\bf B539 } (2002)  227-234.
  [hep-ph/0110078].

\bibitem{B2Vg-NP}
%\cite{Mahmoudi:2009zx}
%\bibitem{Mahmoudi:2009zx}
  F.~Mahmoudi, O.~Stal,
  %``Flavor constraints on the two-Higgs-doublet model with general Yukawa couplings,''
  Phys.\ Rev.\  {\bf D81}, 035016 (2010).
  [arXiv:0907.1791 [hep-ph]];
%\cite{Ahmady:2006yr}
%\bibitem{Ahmady:2006yr}
  M.~R.~Ahmady, F.~Mahmoudi,
  %``Constraints on the mSUGRA parameter space from NLO calculation of isospin asymmetry in B ---> K* gamma,''
  Phys.\ Rev.\  {\bf D75}, 015007 (2007).
  [hep-ph/0608212];
%\cite{Ahmady:2005nc}
%\bibitem{Ahmady:2005nc}
  M.~R.~Ahmady, F.~Chishtie,
  %``Isospin symmetry breaking in B- > K* gamma decay due to an extra generation of vector quarks,''
  Int.\ J.\ Mod.\ Phys.\  {\bf A20}, 6229-6240 (2005).
  [hep-ph/0508105];
%\cite{Xiao:2003vq}
%\bibitem{Xiao:2003vq}
  Z.~-j.~Xiao, C.~Zhuang,
  %``Exclusive B ---> (K*, rho) gamma decays in the general two Higgs doublet models,''
  Eur.\ Phys.\ J.\  {\bf C33}, 349-368 (2004).
  [hep-ph/0310097].

%\cite{Appelquist:1974tg}
\bibitem{Appelquist:1974tg}
  T.~Appelquist, J.~Carazzone,
  %``Infrared Singularities and Massive Fields,''
  Phys.\ Rev.\  {\bf D11}, 2856 (1975).

%\cite{Grzadkowski:2010es}
\bibitem{Grzadkowski:2010es}
  B.~Grzadkowski, M.~Iskrzynski, M.~Misiak, J.~Rosiek,
  %``Dimension-Six Terms in the Standard Model Lagrangian,''
  JHEP {\bf 1010}, 085 (2010).
  [arXiv:1008.4884 [hep-ph]].

%\cite{Buchmuller:1985jz}
\bibitem{Buchmuller:1985jz}
  W.~Buchmuller, D.~Wyler,
  %``Effective Lagrangian Analysis of New Interactions and Flavor Conservation,''
  Nucl.\ Phys.\  {\bf B268}, 621 (1986).

%\cite{AguilarSaavedra:2008zc}
\bibitem{AguilarSaavedra:2008zc}
  J.~A.~Aguilar-Saavedra,
  %``A Minimal set of top anomalous couplings,''
  Nucl.\ Phys.\  {\bf B812}, 181-204 (2009).
  [arXiv:0811.3842 [hep-ph]].

%\cite{Hollik:1998vz}
\bibitem{Hollik:1998vz}
  W.~Hollik, J.~I.~Illana, S.~Rigolin, C.~Schappacher, D.~Stockinger,
  %``Top dipole form-factors and loop induced CP violation in supersymmetry,''
  Nucl.\ Phys.\  {\bf B551}, 3-40 (1999).
  [hep-ph/9812298].

%\cite{Zhang:2008yn}
\bibitem{Zhang:2008yn}
  J.~J.~Zhang, C.~S.~Li, J.~Gao, H.~Zhang, Z.~Li, C.~-P.~Yuan, T.~-C.~Yuan,
  %``Next-to-leading order QCD corrections to the top quark decay via model-independent FCNC couplings,''
  Phys.\ Rev.\ Lett.\  {\bf 102}, 072001 (2009).
  [arXiv:0810.3889 [hep-ph]];
%\cite{Zhang:2010bm}
%\bibitem{Zhang:2010bm}
  %J.~J.~Zhang, C.~S.~Li, J.~Gao, H.~X.~Zhu, C.~-P.~Yuan, T.~-C.~Yuan,
  %``Next-to-leading order QCD corrections to the top quark decay via the Flavor-Changing Neutral-Current operators with
  %mixing
  %effects,''
  Phys.\ Rev.\  {\bf D82}, 073005 (2010).
  [arXiv:1004.0898 [hep-ph]];
%\cite{Zhang:2011gh}
%\bibitem{Zhang:2011gh}
  Y.~Zhang, B.~H.~Li, C.~S.~Li, J.~Gao, H.~X.~Zhu,
  %``Next-to-leading order QCD corrections to the top quark associated with $\gamma$ production via model-independent
  %flavor-changing
  %neutral-current couplings at hadron colliders,''
  [arXiv:1101.5346 [hep-ph]].

%\cite{Drobnak:2010wh}
\bibitem{Drobnak:2010wh}
  J.~Drobnak, S.~Fajfer, J.~F.~Kamenik,
  %``Flavor Changing Neutral Coupling Mediated Radiative Top Quark Decays at Next-to-Leading Order in %QCD,''
  Phys.\ Rev.\ Lett.\  {\bf 104}, 252001 (2010).
  [arXiv:1004.0620 [hep-ph]];
%\cite{Drobnak:2010by}
%\bibitem{Drobnak:2010by}
  %J.~Drobnak, S.~Fajfer, J.~F.~Kamenik,
  %``QCD Corrections to Flavor Changing Neutral Coupling Mediated Rare Top Quark Decays,''
  Phys.\ Rev.\  {\bf D82}, 073016 (2010).
  [arXiv:1007.2551 [hep-ph]].

\bibitem{CKM}
%\cite{Cabibbo:1963yz}
%\bibitem{Cabibbo:1963yz}
  N.~Cabibbo,
  %``Unitary Symmetry and Leptonic Decays,''
  Phys.\ Rev.\ Lett.\  {\bf 10}, 531-533 (1963);
%\cite{Kobayashi:1973fv}
%\bibitem{Kobayashi:1973fv}
  M.~Kobayashi, T.~Maskawa,
  %``CP Violation in the Renormalizable Theory of Weak Interaction,''
  Prog.\ Theor.\ Phys.\  {\bf 49}, 652-657 (1973).

%\cite{Chetyrkin:1996vx}
\bibitem{Chetyrkin:1996vx}
  K.~G.~Chetyrkin, M.~Misiak, M.~Munz,
  %``Weak radiative B meson decay beyond leading logarithms,''
  Phys.\ Lett.\  {\bf B400}, 206-219 (1997).
  [hep-ph/9612313].

\bibitem{WC1}
%\cite{Buchalla:1995vs}
%\bibitem{Buchalla:1995vs}
  G.~Buchalla, A.~J.~Buras, M.~E.~Lautenbacher,
  %``Weak decays beyond leading logarithms,''
  Rev.\ Mod.\ Phys.\  {\bf 68}, 1125-1144 (1996).
  [hep-ph/9512380].
%\cite{Buras:1998raa}
%\bibitem{Buras:1998raa}
  A.~J.~Buras,
  %``Weak Hamiltonian, CP violation and rare decays,''
  [hep-ph/9806471].

\bibitem{WC2}
%\cite{Misiak:2004ew}
%\bibitem{Misiak:2004ew}
  M.~Misiak, M.~Steinhauser,
  %``Three loop matching of the dipole operators for b ---> s gamma and b ---> s g,''
  Nucl.\ Phys.\  {\bf B683}, 277-305 (2004).
  [hep-ph/0401041];
%\cite{Gorbahn:2004my}
%\bibitem{Gorbahn:2004my}
  M.~Gorbahn, U.~Haisch,
  %``Effective Hamiltonian for non-leptonic |Delta F| = 1 decays at NNLO in QCD,''
  Nucl.\ Phys.\  {\bf B713}, 291-332 (2005).
  [hep-ph/0411071];
%\cite{Gorbahn:2005sa}
%\bibitem{Gorbahn:2005sa}
  M.~Gorbahn, U.~Haisch, M.~Misiak,
  %``Three-loop mixing of dipole operators,''
  Phys.\ Rev.\ Lett.\  {\bf 95}, 102004 (2005).
  [hep-ph/0504194];
%\cite{Czakon:2006ss}
%\bibitem{Czakon:2006ss}
  M.~Czakon, U.~Haisch, M.~Misiak,
  %``Four-Loop Anomalous Dimensions for Radiative Flavour-Changing Decays,''
  JHEP {\bf 0703}, 008 (2007).
  [hep-ph/0612329].

%\cite{Feldmann:2002iw}
\bibitem{Feldmann:2002iw}
  T.~Feldmann, J.~Matias,
  %``Forward backward and isospin asymmetry for B ---> K* l+ l- decay in the standard model and in supersymmetry,''
  JHEP {\bf 0301}, 074 (2003).
  [hep-ph/0212158].

%\cite{Li:1990qf}
\bibitem{Li:1990qf}
  C.~S.~Li, R.~J.~Oakes, T.~C.~Yuan,
  %``QCD corrections to $t \to W^{+} b$,''
  Phys.\ Rev.\  {\bf D43}, 3759-3762 (1991).

%\cite{:1900yx}
\bibitem{:1900yx}
  [ CDF and D0 Collaboration ],
  %``Combination of CDF and D0 Results on the Mass of the Top Quark using up to 5.6 $fb^{-1}$ of data,''
  [arXiv:1007.3178 [hep-ex]].

%\cite{Wolfenstein:1983yz}
\bibitem{Wolfenstein:1983yz}
  L.~Wolfenstein,
  %``Parametrization of the Kobayashi-Maskawa Matrix,''
  Phys.\ Rev.\ Lett.\  {\bf 51}, 1945 (1983).

%\cite{Charles:2004jd}
\bibitem{Charles:2004jd}
  J.~Charles {\it et al.} [ CKMfitter Group Collaboration ],
  %``CP violation and the CKM matrix: Assessing the impact of the asymmetric $B$ factories,''
  Eur.\ Phys.\ J.\  {\bf C41}, 1-131 (2005).
  [hep-ph/0406184], and updated at http://ckmfitter.in2p3.fr/.

\bibitem{HPQCD_quarkmass}
%\cite{Davies:2009ih}
%\bibitem{Davies:2009ih}
  C.~T.~H.~Davies, C.~McNeile, K.~Y.~Wong, E.~Follana, R.~Horgan, K.~Hornbostel, G.~P.~Lepage, J.~Shigemitsu {\it et
  al.},
  %``Precise Charm to Strange Mass Ratio and Light Quark Masses from Full Lattice QCD,''
  Phys.\ Rev.\ Lett.\  {\bf 104}, 132003 (2010).
  [arXiv:0910.3102 [hep-ph]];
%\cite{McNeile:2010ji}
%\bibitem{McNeile:2010ji}
  C.~McNeile, C.~T.~H.~Davies, E.~Follana, K.~Hornbostel, G.~P.~Lepage,
  %``High-Precision c and b Masses, and QCD Coupling from Current-Current Correlators in Lattice and Continuum QCD,''
  Phys.\ Rev.\  {\bf D82}, 034512 (2010).
  [arXiv:1004.4285 [hep-lat]].

%\cite{Chetyrkin:2000yt}
\bibitem{Chetyrkin:2000yt}
  K.~G.~Chetyrkin, J.~H.~Kuhn, M.~Steinhauser,
  %``RunDec: A Mathematica package for running and decoupling of the strong coupling and quark masses,''
  Comput.\ Phys.\ Commun.\  {\bf 133}, 43-65 (2000).
  [hep-ph/0004189].

%\cite{Beneke:1998rk}
\bibitem{Beneke:1998rk}
  M.~Beneke,
  %``A Quark mass definition adequate for threshold problems,''
  Phys.\ Lett.\  {\bf B434}, 115-125 (1998).
  [hep-ph/9804241];
%\cite{Beneke:1999fe}
%\bibitem{Beneke:1999fe}
  M.~Beneke, A.~Signer,
  %``The Bottom MS-bar quark mass from sum rules at next-to-next-to-leading order,''
  Phys.\ Lett.\  {\bf B471}, 233-243 (1999).
  [hep-ph/9906475].

%\cite{Laiho:2009eu}
\bibitem{Laiho:2009eu}
  J.~Laiho, E.~Lunghi, R.~S.~Van de Water,
  %``Lattice QCD inputs to the CKM unitarity triangle analysis,''
  Phys.\ Rev.\  {\bf D81}, 034503 (2010).
  [arXiv:0910.2928 [hep-ph]].

\end{thebibliography}
\end{document}